\documentclass[%
 reprint,
superscriptaddress,
 amsmath,amssymb,
 aps,
floatfix,
longbibliography
]{revtex4-2}

\usepackage{graphicx}
\usepackage{dcolumn}
\usepackage{bm}
\usepackage{todonotes}
\usepackage{hyperref}
\usepackage{xspace}

\newcommand{\rrscan}{r$^2$SCAN\xspace}

\newcommand{\tarpbesol}{$\delta\Delta H^{\mathrm{PBEsol}}_{f}$\xspace}

\newcommand{\tarrscan}{$\delta\Delta H^{\mathrm{MLIP}}_{f}$\xspace}

\begin{document}

\preprint{APS/123-QED}

\title{
Correcting DFT formation energies towards experimental accuracy using foundational MLIPs and latent-feature delta-learning
}

\author{Timo Reents}
\affiliation{
PSI Center for Scientific Computing, Theory and Data, Paul Scherrer Institute, 5232 Villigen PSI, Switzerland
}%
\affiliation{
National Centre for Computational Design and Discovery of Novel Materials (MARVEL), Paul Scherrer Institut, 5232 Villigen PSI, Switzerland\\
$^{*}$Corresponding author: giovanni.pizzi@psi.ch
}%

\author{Marnik Bercx} 
\affiliation{
PSI Center for Scientific Computing, Theory and Data, Paul Scherrer Institute, 5232 Villigen PSI, Switzerland
}%
\affiliation{
National Centre for Computational Design and Discovery of Novel Materials (MARVEL), Paul Scherrer Institut, 5232 Villigen PSI, Switzerland\\
$^{*}$Corresponding author: giovanni.pizzi@psi.ch
}%

\author{Giovanni Pizzi$^{*,}$}
\affiliation{
PSI Center for Scientific Computing, Theory and Data, Paul Scherrer Institute, 5232 Villigen PSI, Switzerland
}%
\affiliation{
National Centre for Computational Design and Discovery of Novel Materials (MARVEL), Paul Scherrer Institut, 5232 Villigen PSI, Switzerland\\
$^{*}$Corresponding author: giovanni.pizzi@psi.ch
}%

\date{\today}

\begin{abstract}
\textbf{{\normalsize ABSTRACT}}
\newline

Crystal structure databases curated by high-throughput density functional theory calculations typically serve as the starting point for computational materials discovery efforts. Thermodynamic stability data, such as formation energies and the energy above the convex hull, are important quantities to guide the search for novel materials, enabling filtering for (meta)stable structures. Here, we present the thermodynamic stability of the fully open-source, reproducible, and experimentally focused Materials Cloud three-dimensional crystals database (MC3D). We compare against two other DFT databases, the Open Quantum Materials Database (OQMD) and the Materials Project (MP), as well as against experimental formation enthalpies. We then demonstrate how recent foundational machine learning interatomic potentials (MLIPs) trained at the \rrscan level (specifically, we test PET-OMATPES here) can be leveraged to improve the agreement of formation energies with experiment, reducing the mean absolute error by more than 40\% relative to GGA without requiring any additional DFT calculation. 
Our results validate and extend the established practice of combining PBEsol geometries with meta-GGA energies to the era of foundational MLIPs. Finally, we train classical machine learning models to further correct the formation energies in a delta-learning framework, where we use the information-rich latent features of the foundational MLIP. These models further reduce the mean absolute error below 50~meV/atom, bringing it down to values comparable with the experimental uncertainty itself. Notably, compared to purely compositional features, the latent features---combined with carefully tuned regularization---simultaneously reduce the prediction error and limit the impact of the learned corrections on the relative phase stability.
\end{abstract}

\maketitle

\section*{Introduction}

In recent years, high-throughput workflow engines~\cite{huber_aiida_2020,uhrin_workflows_2021,rosen_jobflow_2024,jain_fireworks_2015,mathew_atomate_2017,ganose_atomate2_2024} have enabled the construction of large databases in the field of computational materials science, including the Materials Project (MP)~\cite{jain_high-throughput_2011,jain_commentary_2013}, the Open Quantum Materials Database (OQMD)~\cite{saal_materials_2013,kirklin_open_2015}, the Automatic FLOW for Materials Discovery (AFLOW)~\cite{curtarolo_aflow_2012,curtarolo_aflowliborg_2012,calderon_aflow_2015}, Alexandria~\cite{schmidt_machine-learning-assisted_2023,schmidt_dataset_2022,schmidt_improving_2024}, and the Materials Cloud three-dimensional crystals database (MC3D)~\cite{huber_mc3d_2026,talirz_materials_2020} discussed here. Typically, these databases are based on density functional theory (DFT)~\cite{hohenberg_inhomogeneous_1964,kohn_self-consistent_1965}, enabling strong predictive power at reasonable computational cost. As the amount of available crystal structure data grows rapidly~\cite{cavignac_ai-driven_2025,parackal_screening_2026}, thermodynamic stability has become an important property for sorting and filtering materials, focusing the search on compounds more likely to exist under experimental conditions~\cite{sun_thermodynamic_2016,meng_computational_2024,wang_predicting_2021}. 
While meta-GGA functionals, such as SCAN~\cite{sun_strongly_2015}, rSCAN~\cite{bartok_regularized_2019} and r$^2$SCAN~\cite{furness_accurate_2020}, are known to produce stability results in better agreement with experimental references~\cite{zhang_efficient_2018,kingsbury_performance_2022,isaacs_performance_2018}, outperforming GGA functionals such as PBE~\cite{perdew_generalized_1996} and PBEsol~\cite{perdew_restoring_2008}, and are increasingly applied in recent high-throughput studies~\cite{schmidt_dataset_2022,kingsbury_flexible_2022,kuner_mp-aloe_2025,malosso_high-quality_2026,kaplan_foundational_2025}, the majority of data in the aforementioned databases is still calculated at the level of the generalized gradient approximation (GGA) due to its computational efficiency, wide adoption in the community, as well as the availability of verified pseudopotentials~\cite{bosoni_how_2024,prandini_precision_2018}, while the latter are still generally missing for meta-GGA functionals. However, GGA formation energies are known to exhibit limited accuracy, e.g. due to the overbinding of diatomic gases~\cite{grindy_approaching_2013} and systematic errors in the description of compounds with localized electrons, e.g. transition metal oxides~\cite{wang_oxidation_2006}.

To improve on the known limitations of GGA formation energies with respect to experimental data, various approaches have been developed. The most common is the correction of elemental reference states, known as Fitted Elemental Reference Energies (FERE)~\cite{stevanovic_correcting_2012}, which addresses overbinding and improves DFT-experiment agreement at the cost of additional empiricism. Different variations of this approach have been adopted across established databases: the OQMD adjusts elemental references only for a subset of chemical elements~\cite{kirklin_open_2015}, while MP uses a slightly different subset with environment-dependent corrections for oxygen~\cite{wang_framework_2021,jain_commentary_2013}, and AFLOW fits corrections based on the local chemical environment of each element~\cite{friedrich_coordination_2019,friedrich_aflow-cce_2024}. In general, empirical corrections represent a trade-off between improved formation energies and additional uncertainty introduced by the corrections themselves. This trade-off becomes particularly relevant when leveraging machine learning (ML) models to predict DFT formation energies~\cite{hautier_accuracy_2012,peterson_materials_2021} or their differences with respect to experimental data~\cite{gong_calibrating_2022}, since ML errors are potentially less systematic and do not benefit from error cancellation to the same extent as DFT~\cite{bartel_critical_2020}.

In this work, we first compare the newly introduced formation energies of the MC3D database with those reported in the established databases OQMD and Materials Project. Since MC3D relies on a different DFT code and computational parameters, this analysis contributes to the comparability of DFT codes and downstream databases, extending existing verification efforts~\cite{hegde_quantifying_2023}. We then address the challenge of improving agreement between DFT calculated formation energies and experiment. We show how existing GGA databases can benefit from foundational machine learning interatomic potentials (fMLIPs) trained at the meta-GGA level, significantly improving the description of thermodynamic stability without requiring additional DFT calculations. Building on this, we show how the information-rich latent features~\cite{gouvea_combining_2026,kim_leveraging_2026} of such fMLIPs can be used to further correct DFT calculated formation energies in a delta-learning framework, reaching a MAE of less than 50~meV/atom with respect to experiment, improving on existing approaches~\cite{gong_calibrating_2022,adhikari_interpretable_2023}. Finally, we examine how regularization, when carefully tuned, can balance an improved accuracy of formation energies with a controlled impact on the relative phase stability.

\begin{figure*}
    \centering
    \includegraphics[width=\linewidth]{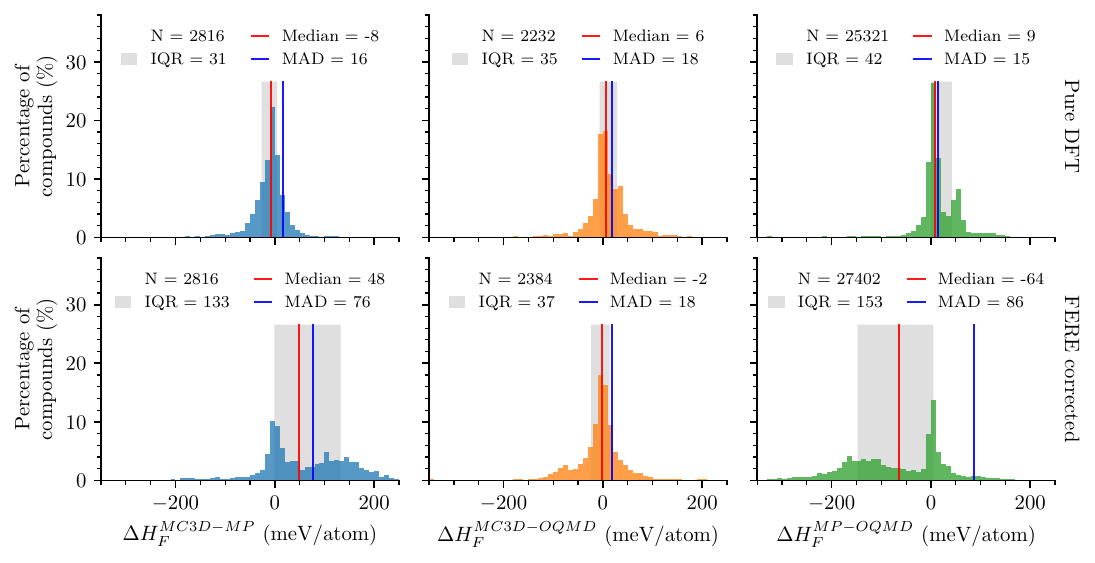}
    \caption{
    \textbf{
    Distribution of  the formation energy differences between different databases: MC3D~\cite{huber_mc3d_2026}, Materials Project~\cite{jain_high-throughput_2011,jain_commentary_2013} and OQMD~\cite{kirklin_open_2015}.}
    \textbf{(upper panels)} The differences of the uncorrected ``pure DFT'' formation energies. \textbf{(lower panels)} The differences across the databases for the empirically corrected formation energies using the FERE approach. Different statistics are shown in the legend: the number of observations $N$, the inter-quartile range $IQR$ ($Q_1 - Q_3$, where $Q_1$ ist the first quartile, equivalent to the 25\% percentile, and $Q_3$ the third quartile, equivalent to the 75\% percentile), the median and the median absolute difference (MAD), whereby the IQR, median and MAD are given in~meV/atom. Only structures with an identical ICSD ID are compared. Moreover, we removed all transition metal compounds containing F or O, since those get a Hubbard $U$ correction in MP which would not be comparable with the MC3D, especially in the pure DFT case.
    }
    \label{fig:database_comp}
\end{figure*}

\section*{Results}

\subsection*{Comparison with existing databases}

\begin{figure*}
    \centering
    \includegraphics[width=\linewidth]{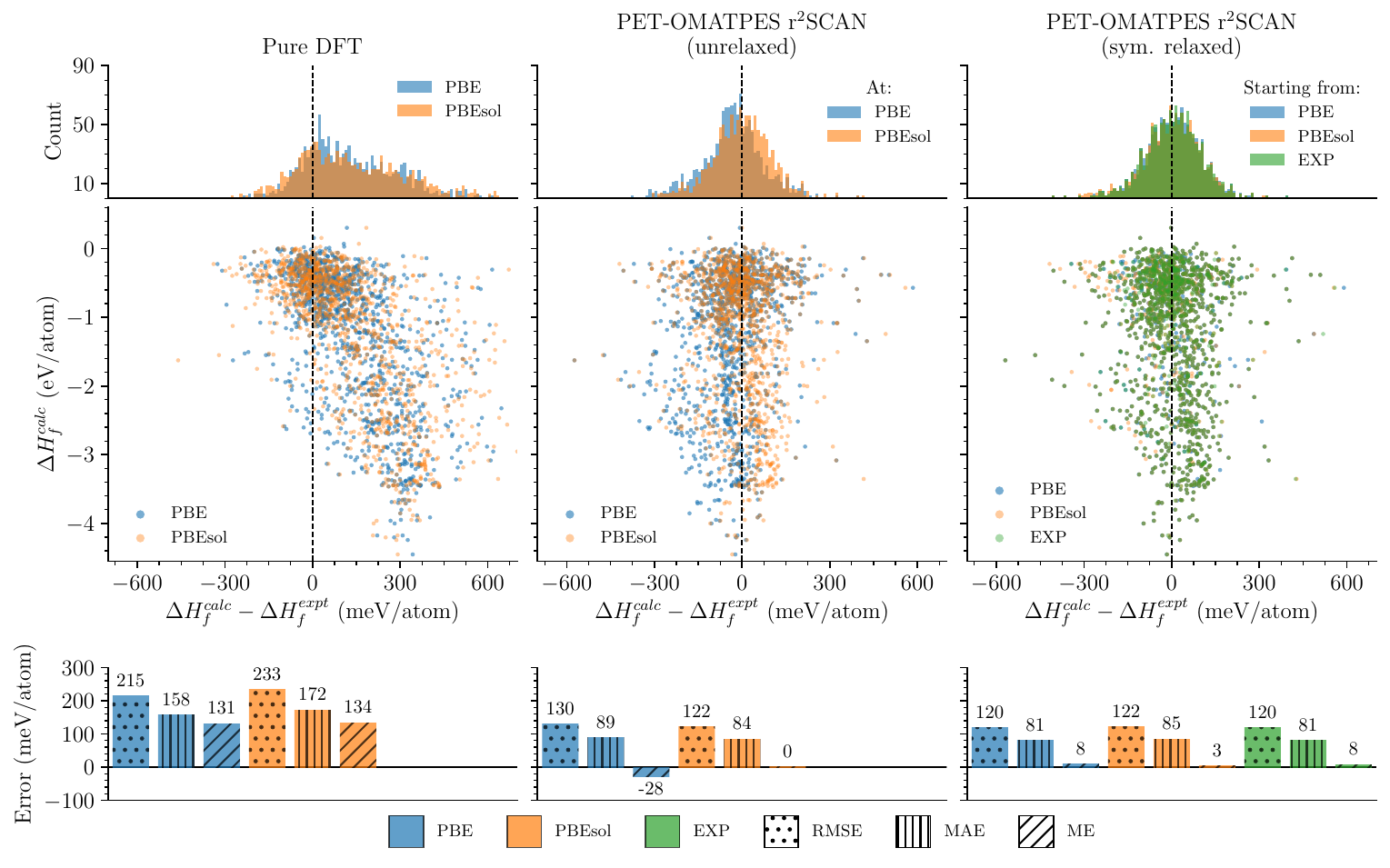}
    \caption{
    \textbf{Comparison of formation energy differences with respect to experiment for several approaches based on PBE, PBEsol and the underlying experimental reference structures EXP.} The different approaches ``calc'' to calculate formation energies are ``Pure DFT'' calculations \textbf{(left column)}, MLIP predicted formation energies evaluated on DFT-relaxed structures ``PET-OMATPES \rrscan (unrelaxed)'' \textbf{(center column)}, and fully MLIP relaxed structures with corresponding MLIP formation energies ``PET-OMATPES \rrscan (sym. relaxed)'' \textbf{(right column)}, using the PET-OMATPES \rrscan MLIP.
    \textbf{(Top row)} Histograms of the per-structure formation energy difference $\Delta H_f^\mathrm{calc} - \Delta H_f^\mathrm{expt}$ at/starting from the PBE relaxed structures (blue), PBEsol relaxed structures (orange) and the experimental source structures EXP (green). EXP is only used for the PET-OMATPES relaxations to investigate to which extent the MLIP relaxations reproduce the ones when starting from the DFT relaxed structures.
    \textbf{(Middle row)} Scatter plots of the calculated formation energy $\Delta H_f^\mathrm{calc}$ against the same difference with respect to experiment, with each point representing one structure. \textbf{(Bottom row)} Summary error metrics of the differences $\Delta H_f^\mathrm{calc} - \Delta H_f^\mathrm{expt}$ with respect to experiment: root mean squared error (RMSE), mean absolute error (MAE), and mean error (ME), for PBE, PBEsol and EXP separately.
    }
    \label{fig:DFT-vs-EXP-functionals}
\end{figure*}

Since we introduce formation energies for MC3D~\cite{huber_mc3d_2026} in this work, we start with comparisons against established databases~\cite{kirklin_open_2015,jain_commentary_2013,horton_accelerated_2025} before addressing the core contributions. 

Fig.~\ref{fig:database_comp} shows pairwise formation energy differences between MP, OQMD, and MC3D (PBE-v1, consistent with the use of the PBE exchange--correlation functional in the other databases). The upper panels compare pure DFT formation energies, the lower panels FERE-corrected~\cite{stevanovic_correcting_2012} ones, i.e., empirically corrected, see the \textbf{Methods} section for more details. While the corrected values are those exposed via public interfaces, the empirical corrections are a major source of inter-database differences. The pure DFT results are therefore particularly informative when comparing across codes and parameter sets, with an inter-quartile range (IQR) of 31--42~meV/atom across all pairs.
The MC3D--MP and MC3D--OQMD distributions are nearly symmetric with one main peak and medians close to zero: MC3D--MP shows a slight excess of compounds in the $-50$ to $0$~meV/atom region, while MC3D--OQMD has a secondary feature near $30$~meV/atom. MP--OQMD, by contrast, shows a pronounced second peak at approximately $50$~meV/atom. As shown in Fig.~\ref{fig:db_comparison_oxygen}, this peak---as well as the skewness in MC3D--OQMD---is driven by oxygen-containing compounds, reflecting differences in the chosen oxygen reference across databases.
A median absolute difference (MAD) of 15--18~meV/atom across all pairs confirms high overall agreement.

Inspecting the corrected formation energies, the MC3D--MP pair now shows a bimodal distribution, in contrast to the unimodal pure DFT case. The MP--OQMD results largely agree with~\cite{hegde_quantifying_2023}, though a more negative median and sharper IQR suggest smaller average deviations, likely reflecting the larger dataset used here. The MP correction scheme was also updated (v2023.11.1 vs.\ v2019.05 in~\cite{hegde_quantifying_2023}), now applying corrections to elemental references as well, bringing it closer to the OQMD approach, though the two databases still correct different sets of elements. The general agreement with Hegde \textit{et al.}~\cite{hegde_quantifying_2023} confirms the validity of our analysis.
The bimodal MC3D--MP distribution arises from the same mechanism identified in~\cite{hegde_quantifying_2023} for MP--OQMD: the oxygen correction scheme in MP, as confirmed by Fig.~\ref{fig:db_comp_diff_el}a,c, with additional contributions from other elements.
The MC3D--OQMD distribution is largely preserved between the pure DFT and FERE-corrected cases, with nearly identical IQR and MAD. The median shifts by $\approx 5$~meV/atom towards more negative values, driven by a larger fraction of compounds below $-50$~meV/atom. These are predominantly Se- and Te-containing compounds (Fig.~\ref{fig:db_comp_diff_el}b): no FERE corrections are fitted for these elements in OQMD, whereas corrections are applied here, lowering their formation energies.
In summary, pure DFT formation energies agree well across databases regardless of code or computational parameters (see SI Fig.~\ref{fig:kpoint_comparison_mc3d_mp} and~\ref{fig:fe_cutoff_corr} and SI section \textbf{Comparison of computational parameters} in general) especially relative to the typical DFT--experiment deviation. The IQR is significantly larger for the corrected energies, underscoring that empirical corrections must be carefully constructed and taken into account when comparing materials across databases.

\subsection*{MLIPs to improve DFT versus experiment}

Having validated the MC3D thermodynamic stability data against other computational databases, we now compare the DFT calculated formation energies against experimental formation enthalpies (see \textbf{Methods} for dataset details). This comparison is practically important, as thermodynamic stability is the most widely used descriptor for filtering materials in computational materials science studies~\cite{parackal_screening_2026,emery_high-throughput_2017,priya_accelerated_2021,woods-robinson_assessing_2018,mueller_evaluation_2011, kirklin_high-throughput_2016,aykol_thermodynamic_2018}, and it is particularly relevant for MC3D, which focuses on experimentally known compounds and is intended to support experimentalists.
The GGA functionals PBE and PBEsol have known limitations in describing thermodynamic stability~\cite{stevanovic_correcting_2012,wang_oxidation_2006,wang_framework_2021,friedrich_aflow-cce_2024}, most prominently the overbinding of molecular oxygen, which leads to overestimated formation energies for oxides~\cite{wang_oxidation_2006}. The FERE corrections~\cite{stevanovic_correcting_2012} introduced in the \textbf{Methods} section can reduce the MAE from approximately 150 to 100~meV/atom (see SI  Fig.~\ref{fig:SI-pbe_sol_expt}), but introduce non-smoothness which is critical for MLIP training~\cite{warford_better_2026}; furthermore, the remaining MAE is still significant.

A straightforward improvement is the adoption of a more advanced functional beyond GGAs, such as the meta-GGA functionals SCAN~\cite{sun_strongly_2015}, rSCAN~\cite{bartok_regularized_2019} or \rrscan~\cite{furness_accurate_2020}, which are known to yield formation energies in better agreement with experiment~\cite{kingsbury_performance_2022, isaacs_performance_2018}. While many databases remain at the GGA level, large meta-GGA datasets start to become available now~\cite{kaplan_foundational_2025,malosso_high-quality_2026,kuner_mp-aloe_2025,kingsbury_flexible_2022,schmidt_dataset_2022}. In the following, we explore how existing GGA databases can benefit from the latest MLIPs trained at the \rrscan level, improving formation energy agreement without rerunning the entire database.

Fig.~\ref{fig:DFT-vs-EXP-functionals} (top left) shows the systematic overestimation by GGA formation energies with respect to experimental values, especially for strongly bound systems~\cite{kingsbury_performance_2022}. The error metrics in Fig.~\ref{fig:DFT-vs-EXP-functionals} (bottom left) confirm the large MAE ($>$170~meV/atom) and a significant positive bias (ME~$\approx$130~meV/atom); PBE and PBEsol perform similarly poorly on this subset. The small difference in the performance of pure PBE and PBEsol is only due to the presence of different elemental references across the two versions of MC3D. When excluding this aspect, both functionals perform almost identically, see SI Fig.~\ref{fig:SI-pbe_sol_expt}. 

As outlined above, adopting meta-GGA functionals represents the logical next step. Instead of recomputing the full database with meta-GGA DFT, we employ MLIPs trained on meta-GGA reference data.
Thanks to recent advances in foundational MLIPs, these models can be directly applied to chemically diverse datasets such as MC3D. Fig.~\ref{fig:DFT-vs-EXP-functionals}b shows that zero-shot formation energies~\cite{yu_systematic_2024} from the PET-OMATPES model~\cite{pozdnyakov_smooth_2024, bigi_pushing_2026}---one of the top-performing models in community benchmarks at the time of publication~\cite{riebesell_framework_2025}, trained on the OMAT~\cite{barroso-luque_open_2024} and MATPES~\cite{kaplan_foundational_2025} datasets at the \rrscan level---reduce the MAE by more than 40\% when evaluated at the DFT (GGA) relaxed geometry.
For PBE, a small residual bias remains (ME~=~$-$28~meV/atom), much improved from the $+$131~meV/atom of pure PBE. For PBEsol, the bias vanishes entirely (ME~=~0~meV/atom). 
Further improvement for PBE is possible by performing full relaxations (constraining the symmetry) with PET-OMATPES (Fig.~\ref{fig:DFT-vs-EXP-functionals}c), which reduces the bias to only 8~meV/atom and the MAE to 81~meV/atom. For PBEsol, the relaxation has little effect: PBEsol was specifically designed for accurate lattice constants~\cite{csonka_assessing_2009} and yields volumes close to those of \rrscan~\cite{kingsbury_performance_2022}, so the higher-level relaxation changes the structure only marginally, consistently with the vanishing bias already observed in the zero-shot results. In contrast, PBE overestimates volumes, which explains the (small) negative bias in Fig.~\ref{fig:DFT-vs-EXP-functionals} (middle column).
We also include the results when directly starting from the experimental structures (underlying the PBEsol results) in the last column of Fig.~\ref{fig:DFT-vs-EXP-functionals}, see EXP, showing that the MLIP relaxation essentially yields the same agreement in terms of formation energies. Notably, Fig.~\ref{SI-fig:r2scan-pbesol-exp-volume} in the SI further shows that the volume change during PET-OMATPES relaxation is on average actually smaller than the DFT PBEsol one, indicating better agreement in terms of volumes with the experimental structures (we note, though, that the DFT workflows used tighter convergence thresholds, which likely causes certain differences). Moreover, while replacing the symmetry-constrained relaxation with an unconstrained one would slightly improve the error for formation energies (see SI Fig.~\ref{SI-fig:DFT-vs-EXP-functionals}), the symmetry-constrained version is more desirable for an unconstrained architecture as it is the case for the PET models~\cite{pozdnyakov_smooth_2024,bigi_pushing_2026}.

In summary, these results show that foundational MLIPs can already serve as better estimators of formation energies relative to experiment than existing GGA databases. Moreover, our results suggest that even structural relaxation could in principle be replaced by fMLIPs, which is an interesting result of this work as itself. However, even though we started from structures whose relaxations reached convergence with PBEsol, approximately 4\% of the calculations using the fMLIP and starting from the experimental structure did not reach convergence, despite using larger convergence thresholds for the MLIP than those adopted in our DFT workflows. Therefore, especially in the context of existing materials databases that already serve properties associated with structures relaxed at the PBEsol level, our recommendation is to compute formation energies combining (PBEsol) DFT-relaxed relaxed geometries followed by zero-shot \rrscan formation energy calculations using the MLIP. This approach extends to foundational MLIPs the established practice in the DFT community of pairing PBEsol geometries with meta-GGA energies~\cite{schmidt_dataset_2022}.

\subsection*{Learning the correction to improve $\Delta H_f^{\rm DFT}$}
\label{ssec:ML}

\begin{figure*}
    \centering
    \includegraphics[width=\linewidth]{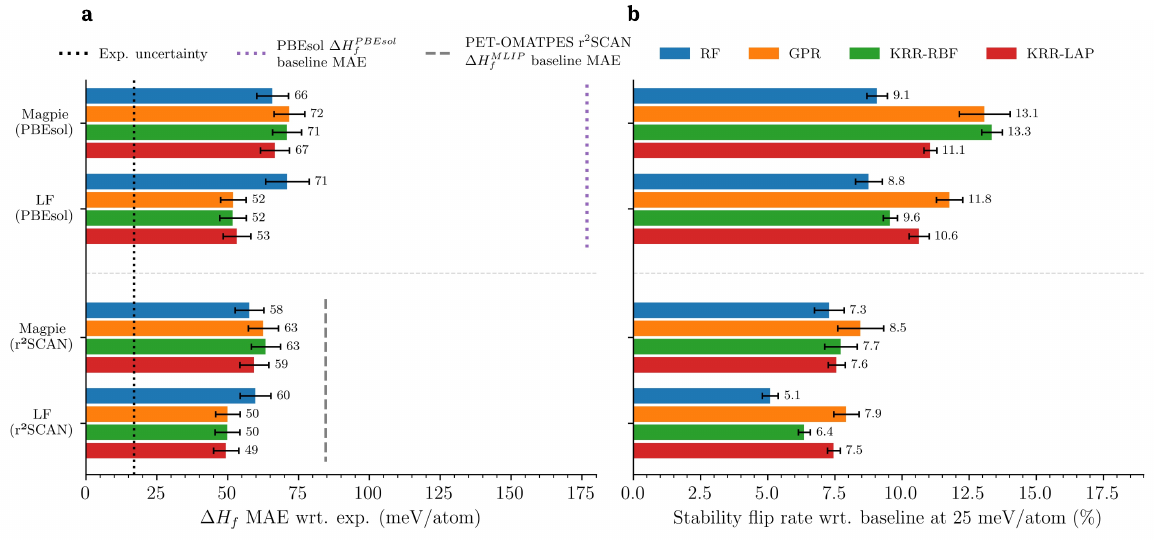}
    \caption{
    \textbf{Performance comparison of different model architectures and feature combinations.}
    \textbf{a} Mean absolute error of predicted formation energies. \textbf{b} Stability flip rate, defined as the fraction of structures whose stability classification differs between the ML prediction and the reference (either PBEsol or PET-OMATPES \rrscan), where we consider as stable materials with an energy above hull smaller than $25$~meV/atom. Results are shown for two feature sets (Magpie and Latent Features from PET-OMATPES (LFs)), both for the PBEsol and \rrscan{} targets, and four model classes: Random Forest, Gaussian Process Regression (GPR), and Kernel Ridge Regression with RBF (KRR-RBF), and Laplacian (KRR-LAP) kernels. Error bars indicate standard deviation across 30 different train-test splits. The dotted purple and dashed gray lines correspond to the baseline MAE with respect to experiment for PBEsol and PET-OMATPES \rrscan{} on the test sets, respectively. The dotted black line indicates the experimental uncertainty.
    }
    \label{fig:ML-model-compare}
\end{figure*}

While the previous section showed that foundational MLIPs already significantly improve agreement with experiment, we now explore how this can be further improved by learning empirical corrections on top.
Beyond simple linear corrections~\cite{wang_oxidation_2006,stevanovic_correcting_2012}, Gong \textit{et al.}~\cite{gong_calibrating_2022} showed that a Random Forest model trained on compositional features (structural features performed worse) reduces the MAE from 95.5 to 61.7~meV/atom on a test set of 229 materials. Notably, the starting point had already incorporated linear corrections, making the initial MAE of 95.5~meV/atom comparably low for PBE. Adhikari \textit{et al.}~\cite{adhikari_interpretable_2023} learned corrections from electronic features for both PBE and SCAN, finding no significant improvement for SCAN.

Building on the studies above, we extend the ML correction approach using the latent features of foundational MLIPs~\cite{chorna_comparing_2026,gouvea_combining_2026,kim_leveraging_2026}. The three models from the \textbf{Methods} section (Random Forest (RF)~\cite{breiman_random_2001}, Kernel Ridge Regression (KRR)~\cite{murphy_machine_2013} with a RBF and Laplacian (LAP) kernel, Gaussian Process Regression (GPR)~\cite{rasmussen_gaussian_2005}) are trained on two targets: (i) \tarpbesol, the difference between PBEsol formation energies and experiment, and (ii) \tarrscan, the difference between zero-shot \rrscan MLIP formation energies (on top of the PBEsol structures) and experiment. For features, we use the Magpie compositional features~\cite{ward_general-purpose_2016} available in \texttt{Matminer}~\cite{ward_matminer_2018} and the PET-OMATPES latent features (LFs), as described in the \textbf{Methods} section.
The Magpie features yield the same correction for each polymorph of a given composition, whereas the LFs assign structure-specific corrections and can therefore affect the relative ordering of phases. The reason for considering here also LFs is that, beyond enabling computationally cheap \rrscan-level formation energy estimates, foundational MLIPs also provide information-rich latent representations learned from millions of crystal structures~\cite{kaplan_foundational_2025,schmidt_improving_2024,horton_accelerated_2025,kuner_mp-aloe_2025}, making their latent space a powerful input for downstream ML corrections. This is further illustrated in SI Fig.~\ref{fig:kpcovr}: a KPCovR map of the LFs shows that the second principal component captures structural similarity, which is not explicitly induced into the model, while the first correlates with the learned correction, validating that the LFs encode correction-relevant information. The map further reveals that the maximum number of unfilled $d$ valence orbitals is a discriminating feature along the correction-correlated axis (see SI section~\textbf{Machine learning of formation energies} for details).

Fig.~\ref{fig:ML-model-compare}a shows the test MAE for each model and feature combination across 30 train-test splits; hyperparameters were selected based on the best average cross-validation MAE across all 30 splits, and each bar reports the average test MAE for models trained with that best hyperparameter set, with error bars indicating the standard deviation across splits.
The \tarrscan{} targets are approximated slightly better than \tarpbesol, with the latter showing a 5--10~meV/atom higher MAE on average, though both are essentially within each other's error bars.
Contrary to what is reported in Gong \textit{et al.}~\cite{gong_calibrating_2022}, we observe that including LF improves performance noticeably for KRR and GPR (but not for RF). Overall, GPR and KRR achieve the best results: a MAE of 52~meV/atom for \tarpbesol{} and 50 and 49~meV/atom for GPR and KRR on \tarrscan, respectively.
Furthermore, in contrast with~\cite{adhikari_interpretable_2023}, we find improved agreement for both the PBEsol and PET-OMATPES \rrscan{} based formation energies with respect to experiment. The relative improvement is larger for PBEsol (178 to 52~meV/atom) than for \rrscan (85 to 49~meV/atom), as indicated by the ratio of the purple (PBEsol) and grey (\rrscan) target lines to the corresponding bars, see Fig.~\ref{fig:ML-model-compare}a. This is also expected because \rrscan energies are already closer to experiment. Notably, both functionals converge to the same final accuracy of around 50~meV/atom, a common lower bound~\cite{kingsbury_flexible_2022} comparable to the experimental uncertainty itself ($\approx$25~meV/atom~\cite{gong_calibrating_2022}, potentially exceeding 70~meV/atom across databases~\cite{kirklin_open_2015}), compared to the approximately 160~meV/atom typical of uncorrected GGA.

\begin{figure}[htb]
    \centering
    \includegraphics[width=\linewidth]{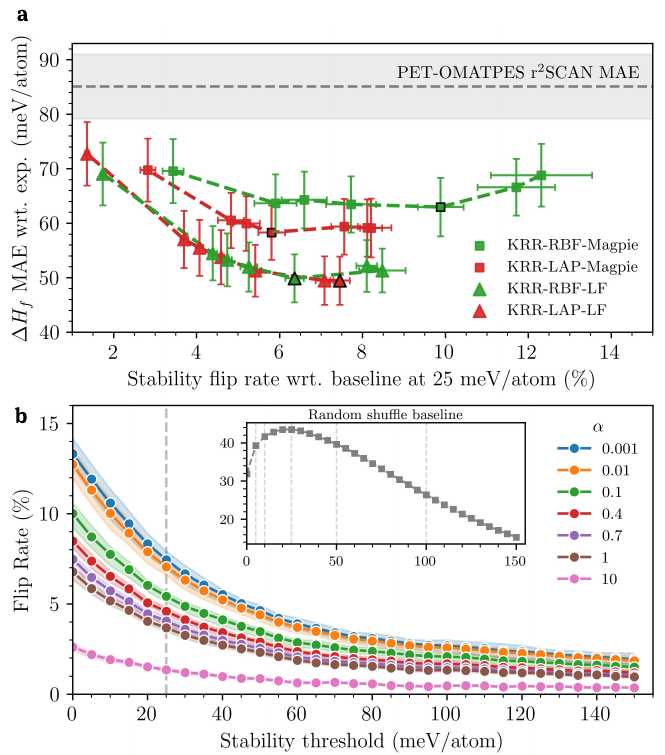}
    \caption{
    \textbf{Regularization impact on the formation energies MAE and stability flip rate.} Both panels are based on the PET-OMATPES \rrscan data.
    \textbf{a} Trade-off between formation energy prediction accuracy and stability reliability for KRR models as a function of regularization strength. Each point corresponds to a specific regularization parameter $\alpha$, with dashed lines connecting points of the same model--feature combination across increasing $\alpha$ values $\alpha \in \{ 0.001, 0.01, 0.1, 0.4, 0.7, 1.0, 10\}$ (generally from left to right). The datapoints highlighted with black edges are the ones that are presented in Fig.~\ref{fig:ML-model-compare}. The x-axis shows the stability flip rate (fraction of structures whose stability classification changes upon correction) and the y-axis the mean test MAE across 30 train-test splits (error bars: standard deviation). Results are shown for four model--feature combinations: KRR with Laplacian and RBF kernels, each paired with Magpie or LF features.
    \textbf{b} Stability flip rate as a function of the stability threshold for the KRR-LAP model with LF features, trained with different regularization parameters $\alpha \in \{ 0.001, 0.01, 0.1, 0.4, 0.7, 1.0, 10\}$. Shaded bands indicate the variance across the 30 train-test splits. Moreover, the inset shows the flip rate expected from randomly sampling the distribution of corrections. The dashed vertical line marks the threshold of 25~meV/atom that is used in panel \textbf{a} and Fig.~\ref{fig:ML-model-compare}b.
    }
    \label{fig:regularization-selection}
\end{figure}

A critical aspect of empirical corrections is their impact on relative phase stability, i.e., whether a material lies on the convex hull or has a finite energy above the hull~\cite{gong_calibrating_2022,adhikari_interpretable_2023}. Fig.~\ref{fig:ML-model-compare}b shows the stability flip rate, i.e., the fraction of structures whose stability classification changes upon correction. Here, we consider a material to be stable if it has an energy above hull smaller than a threshold of $25$~meV/atom; the stability flip rate as a function of the threshold will be discussed later in the context of Fig.~\ref{fig:regularization-selection}b.
We validate the use of PET-OMATPES \rrscan formation energies as the baseline for \tarrscan{} targets against the Alexandria database (see SI Fig.~\ref{fig:SI-alexandria-flip}). Since no meta-GGA DFT version of MC3D currently exists, Alexandria serves as the reference and confirms that the MLIP-estimated stability data agrees well with DFT, validating the use of the PET-OMATPES \rrscan formation energies and energies above hull as the reference for the stability flip rate analysis.

While the MAE differences between models are small (Fig.~\ref{fig:ML-model-compare}a), the flip rates differ more significantly across models.
The flip rates are smaller for the \tarrscan than for the \tarpbesol baseline. However, when the \tarrscan-corrected energies and the resulting energies above hull are evaluated against the PBEsol baseline rather than the PET-OMATPES \rrscan one, the flip rates become comparable. This suggests that a fraction of the flips reflect physically motivated changes already captured by the zero-shot \rrscan formation energies, as confirmed by the analysis based on the  Alexandria database that is shown in SI Fig.~\ref{fig:SI-alexandria-flip} and the discussion of Fig.~\ref{fig:flip-improvement-matrix} later on. This is in fact beneficial: those stability changes are driven by the large-scale PET-OMATPES model trained on millions of structures, rather than by the smaller correction model.

\subsection*{Balance accurate formation energies with distortion of relative phase stability}

Since the \tarrscan{} target yields the best performance, we focus exclusively on it for the remainder of the analysis.
So far, models have been selected based solely on MAE, which favours KRR with either the RBF or LAP kernel. We now examine how regularization, controlled by the $\alpha$ parameter in KRR, mediates the trade-off between formation energy accuracy and distortion of relative phase stability.
Fig.~\ref{fig:regularization-selection}a shows this trade-off for KRR models with Laplacian and RBF kernels, each paired with Magpie or LF features. The RBF and Laplacian kernels perform similarly overall, but the LAP kernel exhibits stronger implicit regularization: the flip rate remains lower even at the smallest $\alpha=0.001$ value.
Notably, LFs enable both a lower MAE and a lower flip rate compared to Magpie features, which is a non-obvious result given that purely compositional features were previously found to be beneficial~\cite{gong_calibrating_2022}.
To provide some additional reference, Fig.~\ref{SI-fig:FE-vs-flip-PBEsol-with-ref} in the SI presents the same analysis as Fig.~\ref{fig:regularization-selection}a but using PBEsol targets instead of the \rrscan ones. Moreover, it also includes a retrained version of the model from Gong~\textit{et al.}~\cite{gong_calibrating_2022} as a reference point. The comparison highlights that even when starting the ML corrections from PBEsol instead of \rrscan, our approach based on the LFs still reduces the MAE on the formation energies and can simultaneously reduce the stability flip rate and its uncertainty.

Fig.~\ref{fig:regularization-selection}b shows the flip rate as a function of the stability threshold. At the strictest threshold (5~meV/atom) with minimal regularization, flip rates can exceed 15\%. Higher regularization reduces them below 10\% or even 5\%, mitigating a challenge highlighted in earlier work~\cite{gong_calibrating_2022}.
Since theoretical studies commonly accept materials within an energy window above the convex hull as stable, acknowledging that experimentally synthesized materials are found at finite energies above the hull~\cite{sun_thermodynamic_2016,bartel_role_2019,parackal_screening_2026}, the impact of corrections is generally less severe at larger thresholds. Common window widths are 25--100~meV/atom~\cite{parackal_screening_2026,emery_high-throughput_2017,priya_accelerated_2021,woods-robinson_assessing_2018,mueller_evaluation_2011,kirklin_high-throughput_2016,aykol_thermodynamic_2018}.
The flip rates converge to around 3--4\% even at large thresholds, suggesting that the remaining cases represent genuinely required stability reclassifications rather than correction artifacts.
Increased regularization also significantly reduces the variance of corrections across data splits, see Fig.~\ref{fig:regularization-selection}b and SI Fig.~\ref{fig:SI-correction-spread}, though few residual structure-specific artifacts persist depending on which structures appear in the training set; a detailed analysis is provided in SI section~\textbf{Variation of corrections across data splits}. Notably, the corrections cluster into only a small number of discrete modes across splits (typically 3--4), with the dominant cluster covering around 80\% of structures, confirming that the variance is structured rather than random (see SI Fig.~\ref{fig:SI-corrections-clustered}). Larger curated experimental reference datasets are the most promising direction to overcome this limitation.

Based on the trade-off in Fig.~\ref{fig:regularization-selection}a and b, we select the KRR-LAP-LF model with $\alpha = 0.1$.
In order to give an intuitive comparison of the magnitude of the the flip rate, we compare it against a random baseline in which the predicted corrections are shuffled across the materials. Here, the assigned predictions are no longer consistent with the actual structures, and very different corrections might be assigned within a given chemical space, which significantly distorts the relative phase stability. In this way, the random sampling serves as a baseline for a model without any systematic error cancellation. This is shown in the inset of Fig.~\ref{fig:regularization-selection}b.
The random sampling leads to a much higher flip rate of up to 45\%, confirming that the learned corrections are far from random~\cite{bartel_critical_2020}. 
This result thus highlights that our model captures meaningful structure--property relationships, enabling them to improve formation-energy predictions while limiting distortion of relative phase stability.

\begin{figure}[th]
    \centering
    \includegraphics[width=\linewidth]{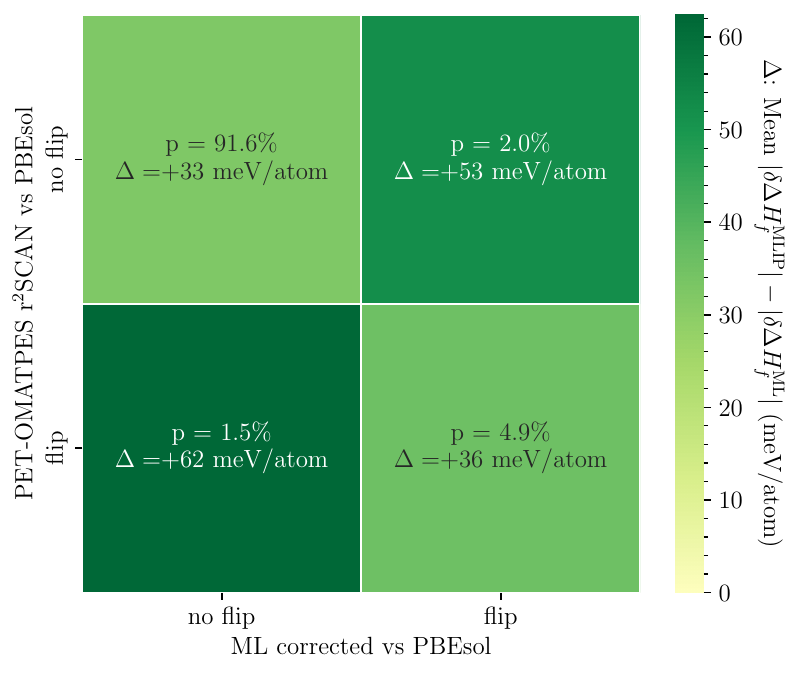}
    \caption{%
    \textbf{Comparison of stability-classification flips induced by zero-shot and ML-corrected PET-OMATPES formation energies.}
    The heatmap shows stability-classification changes relative to the PBEsol baseline for the zero-shot PET-OMATPES \rrscan formation energies, and those with an additional ML correction. The color scale indicates the average difference in agreement with experiment between the two approaches, which is also indicated by the $\Delta$ value within each cell. Positive values correspond to a better agreement of the ML corrected formation energies with respect to experiment compared to the zero-shot PET-OMATPES \rrscan ones, i.e., $|\delta\Delta H_f^\mathrm{ML}|$ is smaller than $|\delta\Delta H_f^\mathrm{MLIP}|$. The fraction $p$ of the structures falling in each category is also indicated. The data is aggregated over the 30 test sets that were analyzed throughout this work.
    }
    \label{fig:flip-improvement-matrix}
\end{figure}

Finally, Fig.~\ref{fig:flip-improvement-matrix} compares the stability flips introduced by the zero-shot PET-OMATPES \rrscan formation energies and by the ML corrected ones, both with respect to the original PBEsol data in MC3D. We first notice that the flips introduced by the PET-OMATPES model and the ML corrected ones agree in around 96.5\% of the cases (sum of the two diagonal blocks). In 1.5\% of the cases, the PET-OMATPES flips are reverted by the ML correction, i.e., the ML corrections on top of PET-OMATPES bring the stability classification back to what was predicted by the PBEsol baseline. Finally, in 2\% of the cases, the PET-OMATPES model does not introduce a flip with respect to PBEsol, but the ML corrections on top do. These results confirm what was already motivated before: the majority of the flips introduced by the ML corrections were already generated by the underlying PET-OMATPES calculation. This motivates again the choice of a two-step correction---first a zero-shot PET-OMATPES \rrscan, then learned ML corrections on top---over the direct delta-learning approach starting from the PBEsol formation energies, as many of the systematic deviations are already covered by the fMLIP, which is physically more grounded. Furthermore, in Fig.~\ref{fig:flip-improvement-matrix} the background color of each block represents the average difference between the absolute value of the error with respect to experiment of the zero-shot PET-OMATPES formation energies \tarrscan{}, and the absolute value of the error after ML correction $\delta\Delta H_f^\mathrm{ML}$. We can observe that, in each category, the ML corrections always further improve (on average) the PET-OMATPES \rrscan{} ones. This aspect is further discussed in Fig.~\ref{SI-fig:heatmap-improvement} and~\ref{SI-fig:heatmap-flips-improvement} in the SI, where we show that the PET-OMATPES formation energies are slightly closer to experiment than the ML corrected ones only in 4\% of the cases.

\subsection*{Comparing with existing approaches}

We finally briefly compare the ML corrected version of the zero-shot PET-OMATPES \rrscan{} formation energies to the well-established FERE approach~\cite{stevanovic_correcting_2012,wang_framework_2021,kirklin_open_2015}, a linear correction scheme commonly applied to GGA databases.

\begin{figure}[htb]
  \centering
  \includegraphics[width=\linewidth]{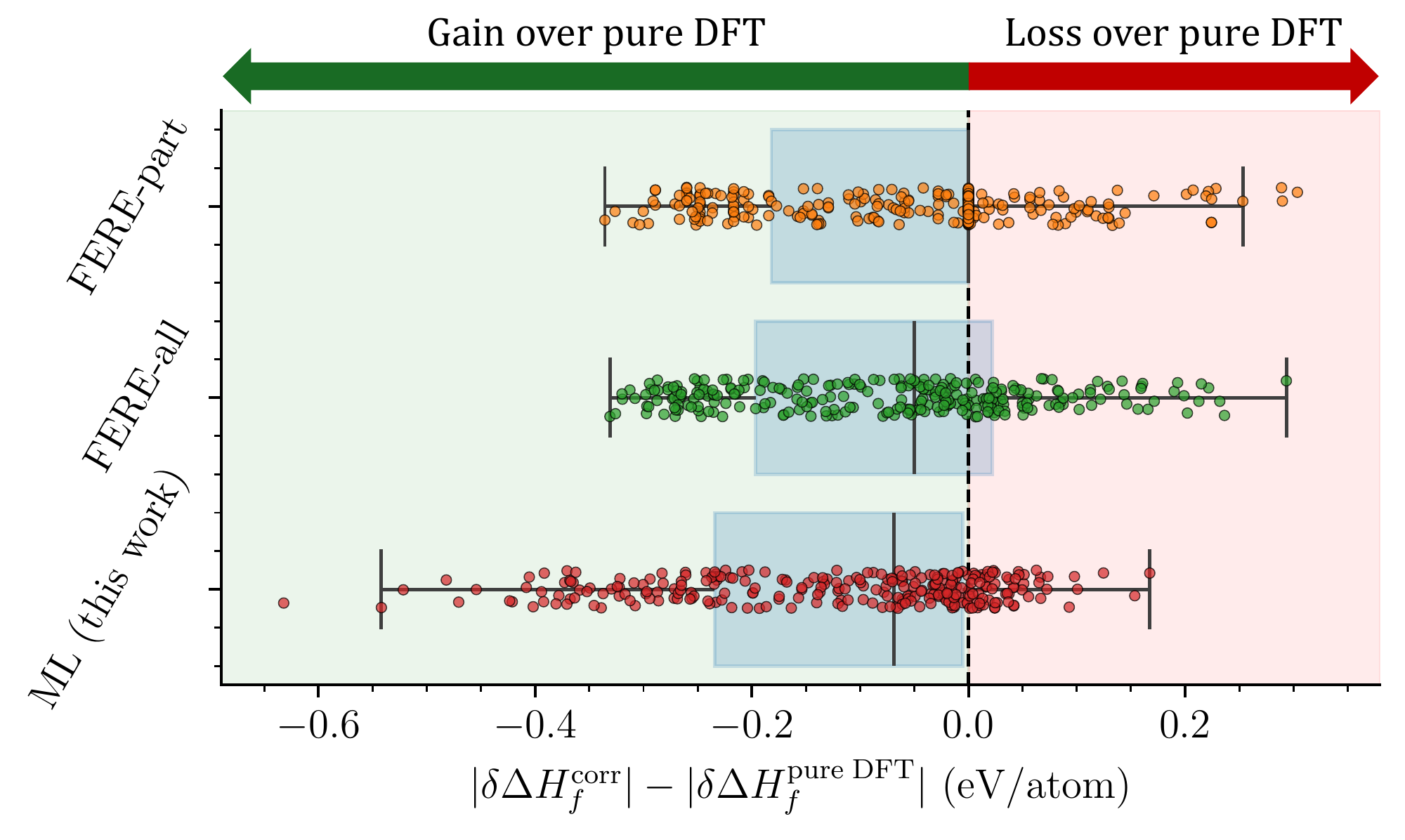}
  \caption{%
  \textbf{Comparison of the ML approach introduced in this work against the established FERE approach.}
  Per-compound change in absolute deviation from experiment, $|\delta\Delta H_f^{\mathrm{corr}}| - |\delta\Delta H_f^{\mathrm{pure\,DFT}}|$, for different correction methods ``corr'': FERE-part, FERE-all, and ML (KRR-LAP-LF, this work). Negative values (green background) indicate improvement over the pure DFT values, positive values (red background) indicate degradation. Box plots show the median (vertical line), IQR (box), and 1.5$\times$IQR whiskers; individual data points are overlaid with vertical jitter.
  }
  \label{fig:ML-FERE-delta-err}
\end{figure}

Fig.~\ref{fig:ML-FERE-delta-err} shows the distribution of $|\delta\Delta H_f^{\mathrm{corr}}| - |\delta\Delta H_f^{\mathrm{pure\,DFT}}|$ for FERE-part (FERE corrections only fitted to a set of elements, see SI section~\textbf{FERE corrections}), FERE-all (FERE corrections fitted to all elements), and ML (our best model KRR-LAP-LF discussed in the previous section), i.e., how the error with respect to experiment of the different correction methods changes compared to the error of pure DFT. While all three methods improve the majority of compounds, they differ notably in how often and how severely they degrade individual predictions. FERE-all worsens 32\% of test compounds (mean degradation 79~meV/atom among those worsened), despite achieving a lower overall MAE than FERE-part; this reflects over-fitting of corrections to sparse elements in the training data. FERE-part worsens 22\% of compounds (mean degradation 107~meV/atom). The ML correction worsens 23\% of compounds, comparable to FERE-part, but with a much smaller mean degradation of only 37~meV/atom. Overall, the ML approach reduces the MAE more substantially than either FERE variant and limits the magnitude of degradation for the minority of compounds it fails to improve. Fig.~\ref{fig:SI-best-fere-ml-error-dist} and~\ref{fig:SI-best-fere-ml-parity} in the SI further visualize the agreement of the different approaches with respect to experiment.

The values in Fig.~\ref{fig:ML-FERE-delta-err} correspond to the data split with the best test MAE of the ML model. However, as shown in SI Fig.~\ref{fig:SI-worst-fere-ml-error-dist}-\ref{fig:SI-ML-FERE-delta-err}, the same trend also holds for the split with the highest test MAE, confirming the results are not an artifact of a specific data split. Moreover, the higher errors are also observed for the pure DFT values, indicating that the reduced performance is driven by certain structures that exhibit much larger errors in general, see SI section~\textbf{Comparison of FERE and ML corrections} for details.

\section*{Discussion}
In summary, we introduced thermodynamic stability data to the MC3D database and validated it against the established databases Materials Project and OQMD. Given that MC3D is based on a different DFT code, different pseudopotentials, and different computational parameters than the VASP-based databases, the good agreement confirms the robustness of the methods adopted in the community, contributing to existing verification efforts~\cite{lejaeghere_reproducibility_2016,bosoni_how_2024,hegde_quantifying_2023}. The comparison also underscores that the main source of inter-database deviations in formation energies are the different empirical correction schemes applied across databases.

The central contribution of this work is the improvement of formation energy agreement with experiment. We show that zero-shot \rrscan formation energies from the foundational MLIP PET-OMATPES already reduce the MAE by more than 40\% relative to GGA, without any additional DFT calculations. This improvement is particularly pronounced when combined with PBEsol-relaxed geometries: since PBEsol lattice constants are already close to those of \rrscan, MLIP relaxation has little effect and the zero-shot evaluation at DFT geometries is nearly optimal. This confirms---and extends to fMLIPs---the established practice of pairing PBEsol geometries with meta-GGA energies~\cite{schmidt_dataset_2022}. Furthermore, we also benchmarked that performing structural relaxations of the original experimental source structures directly with the fMLIP (instead of DFT PBEsol) essentially yields results with the same accuracy in terms of formation energies, and an agreement in terms of experimental volumes that is marginally better than PBEsol.
Going beyond zero-shot MLIP energies, we train classical ML models (KRR, Random Forest, and GPR) in a delta-learning framework to predict the residual between calculated or estimated formation energies and experiment. Using the latent features of PET-OMATPES as input, our best models achieve a MAE below 50~meV/atom  (and potentially even less, depending on the data split) both for the PBEsol and \rrscan targets, becoming comparable to the experimental uncertainty itself. Notably, the latent features simultaneously reduce the MAE and the stability flip rate compared to purely compositional features. We further show that carefully tuned regularization controls the impact of ML corrections on relative phase stability.
Finally, we validate the ML corrections against established FERE empirical corrections, confirming that the ML approach not only reduces the average deviation from experiment more substantially, but also limits the magnitude of degradation for the minority of compounds it fails to improve.

\section*{Methods}

\subsection*{Calculation details}

In this work, we introduce formation energies to the Materials Cloud three-dimensional crystals database~\cite{huber_mc3d_2026}, a fully reproducible and open-source DFT curated database of three-dimensional inorganic crystal structures with a particular focus on experimentally known compounds. Here, we focus on the versions PBE-v1 and PBEsol-v1 of MC3D.

All DFT calculations are performed using SIRIUS-enabled~\cite{sirius} Quantum ESPRESSO version 7.1~\cite{giannozzi_quantum_2009, giannozzi_advanced_2017} together with the PBE~\cite{perdew_generalized_1996} and PBEsol~\cite{perdew_restoring_2008} pseudopotentials (SSSP PBE Efficiency v1.1.0 and SSSP PBEsol Efficiency v1.1.0, respectively) from the SSSP library~\cite{prandini_precision_2018}.
The current MC3D~\cite{huber_mc3d_2026} is based on standard GGA functionals without Hubbard corrections (the absence of a Hubbard U correction can actually be beneficial when training MLIPs, see~\cite{warford_better_2026}).
A k-point spacing of 0.15~\AA$^{-1}$ is used to sample the Brillouin zone following the default protocol of the AiiDA Quantum ESPRESSO plugin~\cite{aiida_qe}. Starting from experimental references, structural parameters and atomic positions are relaxed until forces and energies converge below $10^{-4}$~Ry/bohr and $10^{-5}$~Ry/atom, respectively.
We refer the reader to the original MC3D publication~\cite{huber_mc3d_2026} for details on the curation process, and to Nascimento~\textit{et~al.}~\cite{deMirandaNascimento2026} for the selection of computational parameters.
Formation energies are calculated from the converged relaxation results.

\subsection*{Reference data}
\label{ssec:data}

\paragraph{Experimental references}

To assess the performance of the DFT calculated formation energies, we use the experimental references collected by Wang \textit{et al.}~\cite{wang_framework_2021} and published in the matminer package~\cite{ward_matminer_2018} as the \textit{expt\_formation\_enthalpy\_kingsbury} dataset~\cite{wang_framework_2021,Kim2017,kim_meschel_nash_chen_2017,Kubaschewski1993,NIST,RZYMAN2000309,CRC2007,Grindy2013}. To increase the amount and diversity of experimental reference data, which is typically the limiting factor in data-driven studies, we combine this with the experimental references used by Kirklin \textit{et al.}~\cite{kirklin_open_2015}, yielding 2726 unique compounds in total.

For our analyses, we removed compounds satisfying at least one of the following criteria, balancing data quality against dataset size:

\begin{itemize}
    \item The compound is an elemental phase \mbox{(new size: 2649)}
    \item The experimental uncertainty estimate exceeds 10\%. If no uncertainty is reported, the compound is retained. (new size: 2386)
    \item The DFT formation energy (from OQMD) and experimental reference differ by more than 0.5~eV/atom, removing only a few outliers. (new size: 2373)
    \item The compound appears in both datasets and the two sources differ by more than 150~meV/atom. \mbox{(new size: 2356)}
\end{itemize}

Since MC3D does not contain all compounds, merging with the experimental references yields 1552 (PBE) and 1384 (PBEsol) compounds, of which 1297 overlap. 
When merging calculated data with experimental references, spacegroup information is used where available; otherwise the lowest-energy polymorph at a given composition is selected.

\paragraph{Reference computational databases}

In addition to the comparison with experimental data, the MC3D formation energies are compared against the established high-throughput databases Materials Project~\cite{jain_high-throughput_2011,jain_commentary_2013} and Open Quantum Materials Database~\cite{kirklin_open_2015}. We query the OQMD v1.5 (locally hosted) and MP v2023.11.1, and match structures by their ICSD ID. Compounds without a matching ID are excluded, and no additional structure-based duplicate detection is applied.
MP and OQMD rely on the VASP DFT code~\cite{kresse_efficient_1996, kresse_efficiency_1996}, in contrast to MC3D which uses SIRIUS-enabled Quantum ESPRESSO. This involves differences in pseudopotentials, which play a central role in the precision of the results. Differences in computational parameters across databases are further discussed in SI section~\textbf{Comparison of computational parameters}.

\subsection*{DFT formation energies}
To estimate the thermodynamic stability of materials in MC3D, we calculate the formation energy
\begin{equation*}
    \Delta H_f(A^{1}_{n_1}\dots A^{N}_{n_N}) = E_{tot}(A^{1}_{n_1}\dots A^{N}_{n_N}) - \sum_{i=1}^N n_i \mu(A^i),
\end{equation*}
where $E_{tot}$ is the total energy of the compound consisting of chemical elements $A^1, A^2,\dots A^N$, and $n_i$ is the number of sites occupied by element $A^i$ in the unit cell. The chemical potential $\mu(A^i)$ is given by the DFT total energy per atom of the most stable elemental polymorph in MC3D. This definition is strictly valid at 0~K, whereas experimental formation energies are typically measured under standard conditions. The resulting finite-temperature error is typically of the order of a few tens of meV/atom, which is below the MAE between DFT and experiment, and is therefore neglected~\cite{wang_framework_2021, kirklin_open_2015}.
As discussed in the introduction, GGA formation energies have known systematic limitations. We adopt the FERE corrections~\cite{stevanovic_correcting_2012} in the database comparison and as a baseline for the ML corrections; details on the fitting procedure and comparison with ML corrections are provided in SI sections~\textbf{FERE corrections} and~\textbf{Comparison of FERE and ML corrections}, respectively. Formation energies without empirical corrections are labelled \textit{Pure DFT} throughout this work.

\subsection*{Machine learning models and training}
\label{ssec:ML_training}

In the analysis related to the ML correction of formation energies, we train the following ML models as provided by \texttt{scikit-learn}: \texttt{Random Forest} (RF), \texttt{Gaussian Process Regression} (GPR) and \texttt{Kernel Ridge Regression} (KRR). We test a \texttt{laplacian} (LAP) and \texttt{RBF} kernel for the latter. The adopted framework follows a delta-learning approach, i.e., the model tries to learn the difference between the calculated formation energy and experiment, which can be applied as a correction after successful training.

The dataset of 1384 structures with experimental references is divided into a 80/20 train-test split. In order to estimate the uncertainty, particularly relevant for such a small dataset where individual materials might significantly alter the learned corrections, this split is generated across 30 different seeds. For each of the train-test splits, we perform 5-fold cross validation on the training set to find the best hyperparameters for each model. 
Furthermore, the best set of hyperparameters is used to calculate the test error for each seed. A stratified split was used for the hyperparameter optimization and train-test splits, based on the number of elements per structure, to ensure that each split covers the same (representative) fraction of unaries, binaries etc.

Two feature sets are used: (i) purely compositional features from \texttt{Magpie}~\cite{ward_general-purpose_2016} and (ii) structural features derived from the latent space of the PET-OMATPES model~\cite{pozdnyakov_smooth_2024,bigi_pushing_2026}, following the extraction approach of Chorna~\textit{et al.}~\cite{chorna_comparing_2026}. fMLIPs learn information-rich representations when trained to predict energies and forces across a broad chemical range, making their latent space a natural source of structural descriptors. We evaluate both last-layer~(LL) features and backbone~(BB) features.
The latter are extracted after the message-passing iterations and before the final multilayer perceptron (MLP) layers, that are typically used in MLIPs as the so-called readout layers. As BB features consistently outperform LL features, they are used throughout and referred to as latent features (LF) in the current work.

\section*{Acknowledgments}
This research was supported by the NCCR MARVEL, a National Centre of Competence in Research, funded by the Swiss National Science Foundation (grant number 205602).
This work was supported by a grant from the Swiss National Supercomputing Centre (CSCS) on the Swiss share of the LUMI system under project 465000416. 
Moreover, we acknowledge the use of the Merlin7 cluster run by the Paul Scherrer Institute PSI. M.B. and G.P. acknowledge financial support by the SwissTwins project, funded by the Swiss State Secretariat for Education, Research and Innovation (SERI).

We thank Prof. Nicola Marzari for guidance and interesting discussions of the results.

\section*{Data availability}
The data will be made publicly available on the Materials Cloud Archive~\cite{talirz_materials_2020} at~\url{https://doi.org/10.24435/materialscloud:yb-gw} upon publication.

\section*{Code availability}
The code will be made publicly available upon publication.

\section*{Author contributions}
G.P. and M.B. conceived the project. T.R. contributed to the planning of the project, implemented the code, and performed the calculations and analyses, under the supervision of G.P.
T.R. wrote the initial version of the manuscript with input from M.B. and G.P. 
All authors discussed the results and contributed to editing and the final version of the manuscript.

\section*{Competing interests}
The authors declare no competing financial or non-financial interests.

\bibliographystyle{apsrev4-2}
\bibliography{
formation-energies
}

\clearpage
\pagebreak

\onecolumngrid
\renewcommand{\figurename}{SUPPLEMENTARY FIG.}
\setcounter{figure}{0}
\renewcommand{\thefigure}{S\arabic{figure}}
\renewcommand{\tablename}{SUPPLEMENTARY TABLE}
\setcounter{table}{0}
\renewcommand{\thetable}{S\arabic{table}}
\setcounter{equation}{0}
\renewcommand{\theequation}{S\arabic{equation}}

\renewcommand{\thesection}{S\arabic{section}}
\setcounter{section}{0}

\renewcommand{\theHfigure}{S\arabic{figure}}
\renewcommand{\theHtable}{S\arabic{table}}
\renewcommand{\theHequation}{S\arabic{equation}}
\renewcommand{\theHsection}{S\arabic{section}}

\begin{center}
\textbf{\large Supplemental Materials: Correcting DFT formation energies towards experimental accuracy using foundational MLIPs and latent-feature delta-learning}
\end{center}

\section{Database comparison}

\subsection{Formation energies -- O containing compounds}

\begin{figure}[h!]
    \centering
    \includegraphics[width=\linewidth]{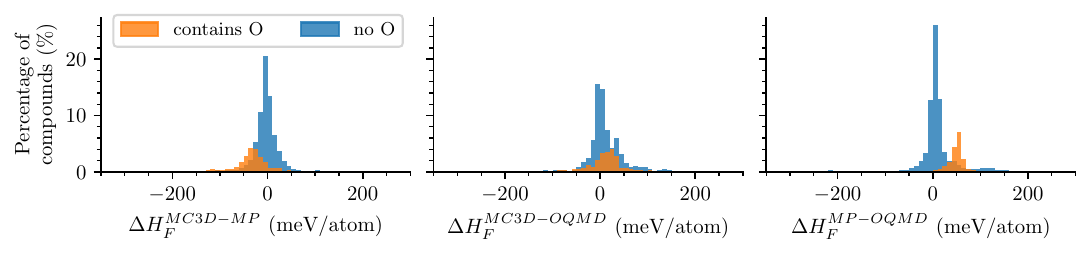}
    \caption{Distribution of formation energy differences between the databases MC3D, MP and OQMD. The color code distinguishes compounds containing oxygen and those that do not contain oxygen. The sum of the two histograms yields the bimodal distribution shown in Fig.~\ref{fig:database_comp} in the main text.}
    \label{fig:db_comparison_oxygen}
\end{figure}

Fig.~\ref{fig:db_comparison_oxygen} presents again the DB comparison of the pure DFT results, see main text Fig.~\ref{fig:database_comp}, highlighting compounds that contain oxygen. One can observe that the bimodal distributions shown in the main text are mostly caused by the oxygen containing compounds.

\subsection{Elements inducing differences between databases}

\begin{figure}[htb!]
    \centering
    \includegraphics[width=.9\linewidth]{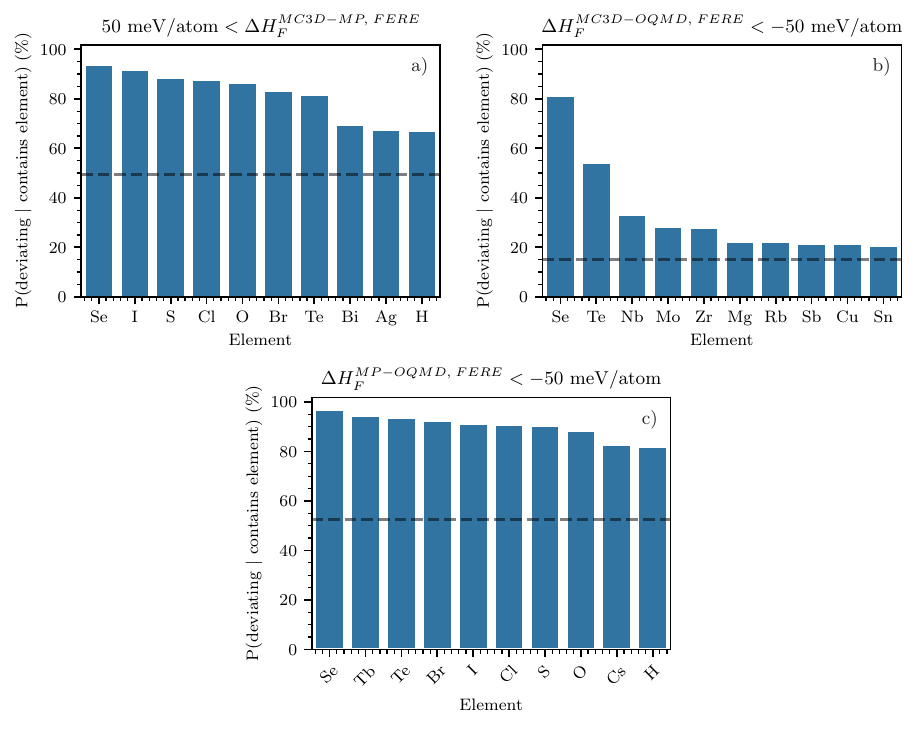}
    \caption{Conditional probability \mbox{$P(\Delta H_F^{D_1-D_2} \in I \;|\;        A^{i}\; \mathrm{in} \; A^{1}_{n_1}\dots A^{N}_{n_N})$} that the formation energy difference between two databases $D_1$ and $D_2$ falls within a certain energy range $I$, given  that a compound $A^{1}_{n_1}\dots A^{N}_{n_N}$ contains a certain element $A^{i}$. The dashed black line indicates the probability independent of the element. We only include elements which have an overall percentage of more than 2\%. \textbf{a} \mbox{$50\mathrm{meV/atom} < \Delta H_F^{MC3D-MP,\;FERE}$}, \mbox{\textbf{b} $\Delta H_F^{MC3D-OQMD,\;FERE} < -50\mathrm{meV/atom}$},
     \textbf{c} \mbox{$\Delta H_F^{MP-OQMD,\;FERE} < -50\mathrm{meV/atom}$}.
    }
    \label{fig:db_comp_diff_el}
\end{figure}

Fig.~\ref{fig:db_comp_diff_el} presents the conditional probability of finding deviating formation energies in databases (within a specified interval), given that the compound contains a certain element.

\subsection{
Comparison of computational parameters
}
\label{SI-ssec:DB-comp-parameters}

One of the key differences in the computational parameters adopted by different high-throughput databases are the k-point sampling to integrate the Brillouin zone and the cutoffs for the wavefunctions and charge density.

\subsubsection{k-points}

\begin{figure}[h!]
    \centering
    \includegraphics{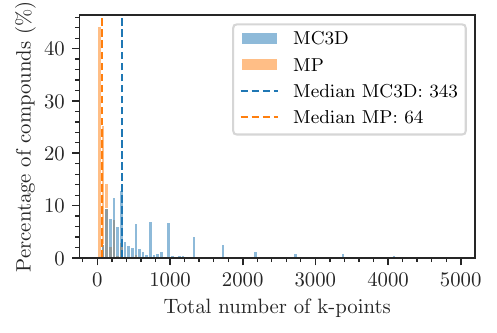}
    \caption{Comparison of the total number of k-points, calculated as the product of the components of the Monkhorst-Pack mesh. The color distinguishes between MC3D and MP, and the dashed lines indicate the median values. Moreover, both distributions are normalized separately, i.e. they both sum up to 100\%.}
    \label{fig:kpoint_comparison_mc3d_mp}
\end{figure}

OQMD uses a k-point sampling of 6000 kpoints per reciprocal atom, whereas Materials Project adopts a sampling of 1000 k-points per reciprocal atom. However, MC3D follows a slightly different approach by defining a minimum distance of \mbox{0.15 \AA$^{-1}$} between subsequent k-points along each reciprocal space direction to determine the k-mesh. Fig.~\ref{fig:kpoint_comparison_mc3d_mp} shows how the k-meshes generated with the MC3D and Materials Project parameters compare across a set of crystal structures. In general, the chosen parameters in the MC3D tend to produce denser k-grids. Moreover, the density of k-points in MC3D was not only chosen to produce dense grids, but carefully tuned to yield optimal results in combination with the chosen smearing~\cite{deMirandaNascimento2026}.

For the comparison of the MC3D and the Materials Project with respect to the adopted k-grid, we query the structures and computational details from the Materials Project. Afterwards, we take these input structures and generate the k-mesh using the MC3D setup. The distribution of the total number of kpoints is visualized in Fig.~\ref{fig:kpoint_comparison_mc3d_mp}

\subsubsection{Plane-wave cutoffs}

Regarding the cutoffs for the wavefunctions and charge density, Materials Project and OQMD use a constant cutoff that is chosen to be 25\% larger than the highest recommended one over all pseudo potentials. In contrast to these constant cutoffs, the MC3D follows the SSSP \cite{prandini_precision_2018} protocol where a set of recommended cutoffs for each element were established. For materials composed of different elements, the wavefunction ($E_{cut}^{wfc}$) and charge density ($E_{cut}^{rho}$) cutoff is chosen as the maximum of each cutoff type across all elements present in the compound, e.g. for MgZnAg$_2$:

\begin{equation}
\begin{split}
    E_{cut}^{wfc}(\mathrm{MgZnAg_2}) &= \max{\left\{E_{cut}^{wfc}(\mathrm{Mg}), E_{cut}^{wfc}(\mathrm{Zn}), E_{cut}^{wfc}(\mathrm{Ag})\right\}}  \\
    E_{cut}^{rho}(\mathrm{MgZnAg_2}) &= \max{\left\{E_{cut}^{rho}(\mathrm{Mg}), E_{cut}^{rho}(\mathrm{Zn}), E_{cut}^{rho}(\mathrm{Ag})\right\}}
\end{split}
\end{equation}

While this approach is very reasonable in general, and imposes a good balance between computational costs and convergence of the results, one needs to carefully check the convergence behavior of relative properties involving values that were calculated with different computational parameters. This is the case for formation energies for which the energy difference of a material and its elemental references is calculated. Revisiting the example of MgZnAg$_2$ mentioned before, one realizes that not all parts that are involved in the calculation of the formation energy are calculated at the same cutoffs. Since the compound cutoffs are chosen as the maximum among the elemental cutoffs, there is typically at least one elemental reference that was calculated with a different cutoff combination. Table~\ref{tab:SI-cutoff-example} presents the cutoffs of Mg, Zn and Ag as well as the resulting cutoffs for different compounds consisting of (all or a subset of) these elements. It is immediately clear that, for this example but generally in several cases, the elemental references and the various compounds are not calculated at the same cutoffs.

\begin{table}[h]
    \centering
    \begin{tabular}{lcc}
    \hline
     & $E_{cut}^{wfc}$ (Ry) & $E_{cut}^{rho}$ (Ry) \\
    \hline
    Mg & 30 & 240 \\
    Zn & 40 & 320 \\
    Ag & 50 & 200 \\
    \hline
    ZnAg    & 50 (Ag) & 320 (Zn) \\
    MgZn    & 40 (Zn) & 320 (Zn) \\
    MgAg    & 50 (Ag) & 240 (Mg) \\
    MgZnAg$_2$ & 50 (Ag) & 320 (Zn) \\
    \hline
    \end{tabular}
    \caption{
    \textbf{Upper part:} $E_{cut}^{wfc}$ and $E_{cut}^{rho}$ for elements Mg, Zn and Ag, as specified in the SSSP PBE Efficiency v1.3.0~\cite{prandini_precision_2018} library.
    \textbf{Lower part:}
    $E_{cut}^{wfc}$ and $E_{cut}^{rho}$ for different materials consisting of the elements Mg, Zn and Ag. In addition to the cutoffs, each cell presents in brackets the element that has determined the cutoff, i.e., the element that exhibits the largest cutoff among the elements that are present in the compound.}
    \label{tab:SI-cutoff-example}
\end{table}

To this end, we generate all possible cutoff combinations that occur across the MC3D database and recalculate the required elemental references to ensure that we have total energy values for all materials and their corresponding elemental references that have been calculated with a consistent cutoff value. The obtained energy values with the adjusted cutoffs are applied to the final formation energies as another type of correction, similar to the FERE approach discussed in the main text.

\begin{figure}[h]
    \centering
    \includegraphics[width=0.5\linewidth]{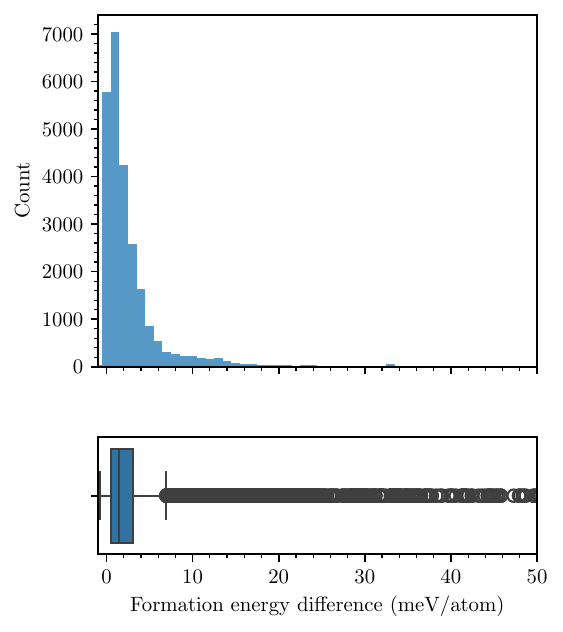}
    \caption{Formation energy difference for all compounds in the MC3D PBE-v1 without and after applying the cutoff corrections. $\Delta H_F^{def}$ and $\Delta H_F^{adj}$ represent the formation energy based on the SSSP default cutoffs and the adjusted cutoffs, respectively.}
    \label{fig:fe_cutoff_corr}
\end{figure}

Finally, we want to note that while it is reasonable from a theoretical perspective and based on the previously presented results to use consistent cutoffs when evaluating formation energies, the adjustment of the cutoffs does not improve the agreement between the DFT formation energies and the experimental references. The difference might be indeed reduced for certain compounds, however on average the MAE remains nearly constant. Moreover, when applying a FERE correction, the adjustment of the cutoffs is not relevant, as the differences are anyway absorbed and incorporated by the FERE corrections themselves. However, in case of the FERE-part formation energies, and especially when dealing with the pure DFT ones, the adjustment of the cutoffs has an effect.
Fig.~\ref{fig:fe_cutoff_corr} shows the distribution of the differences between the formation energies including the cutoff corrections and the ones that are calculated with the default SSSP cutoffs. As outlined above, the impact is relatively small and the majority of compounds exhibits differences below 10 meV/atom. Nonetheless, a set of outliers can be observed, e.g. compounds containing W. Careful analysis showed that this is due to the W cutoffs in the efficiency version of SSSP not being large enough. Indeed, while the tested properties~\cite{prandini_precision_2018} in the SSSP protocol are converged, the total energy still varies significantly when increasing the cutoff. Since the total energy of the elemental reference is the property that is relevant for the calculation of the formation energies, this affects most of the W-containing compounds.

\clearpage

\section{FERE corrections}
\label{SI-sec:fere-corrections}

\paragraph*{Fitting of elemental references}
\label{sssec:FERE}

In this work, we adopt the fitting of the elemental references~\cite{stevanovic_correcting_2012} based on two different sets: all elements that we call \textit{FERE-all} and a subset consisting of \textbf{H, N, O, F, Cl, Br, I, S, Se, Te, P, Sn, Sb} that we call \textit{FERE-part}. The final set of elements used for FERE-part was determined by comparing different elemental sets. We combined the elements that are used in case of OQMD and MP (see~\cite{kirklin_open_2015} and~\cite{hegde_quantifying_2023}) and removed those elements that would only receive a small correction. Moreover, we also compared this approach of applying the corrections to all compounds containing those elements with the conditional one adopted in MP, i.e. only applying the corrections to compounds for which a certain element is contained as an anion. To limit the extent of empiricism, we also tested the approach of only applying the FERE corrections to compounds for which the corresponding element is present as an anion~\cite{wang_framework_2021}. Nonetheless, the performance in terms of MAE differs only by 2 meV/atom between the different methods.
\\
To obtain the corrections applied to the elemental references, the following least-squares problem is solved via linear regression
\begin{equation}
    Ax=b,
\end{equation}
whereby the columns of $A$ correspond to the elements that should be corrected and each row contains the relative amount of the corresponding element in a certain compound. Moreover, $b$ contains the difference between pure DFT and experimental formation energies for each compound. As previous work has shown that the FERE approach is generally stable with respect to overfitting~\cite{isaacs_performance_2018}, the full dataset of 1511 compounds (after removing compounds with an energy above hull larger than 100~meV/atom~\cite{wang_framework_2021}) is used to determine the corrections. The filtering of highly unstable structures (according to PBE) is performed to avoid fitting to outliers.
The resulting corrections per element are presented in table \ref{tab:FERE corrections}. Comparing this to the values obtained by the other high-throughput databases, we observe a relatively high agreement with the ones from OQMD and certain deviations compared to MP.

\begin{table}[h]
\caption{FERE corrections for the MC3D (PBE version), MP and OQMD databases. MP fits different corrections for oxygen based on the local environment, therefore, multiple values are listed for oxygen. All values are presented in eV/atom.}
\label{tab:FERE corrections}

\begin{tabular}{c|c|c|c}
 Element & MC3D FERE correction & MP FERE correction & OQMD FERE correction\\
\colrule
H & $-0.121$ & $-0.179$ & $0.067$ \\
N & $-0.264$ & $-0.361$ & $-0.113$ \\
O & $-0.357$ & ($-0.687$, $-0.465$, $-0.161$) & $-0.359$ \\
F & $-0.387$ & $-0.462$ & $-0.237$ \\
Cl & $-0.319$ & $-0.614$ & $-0.355$\\
Br & $-0.274$ & $-0.534$ & $-0.289$\\
I & $-0.133$ & $-0.379$ & $-0.165$ \\
S & $-0.290$ & $-0.503$ & $-0.275$\\
Se & $-0.172$ & $-0.472$ & - \\
Te & $-0.196$ & $-0.422$ & - \\
P & $-0.301$  & -& $-0.244$ \\
Sn & $-0.254$ & - & $-0.112$\\
Sb & $-0.135$ & $-0.192$ & -\\
\end{tabular}
\end{table}

\subsubsection{PBE vs. PBEsol after FERE corrections}

\begin{figure}[h]
    \centering
    \includegraphics{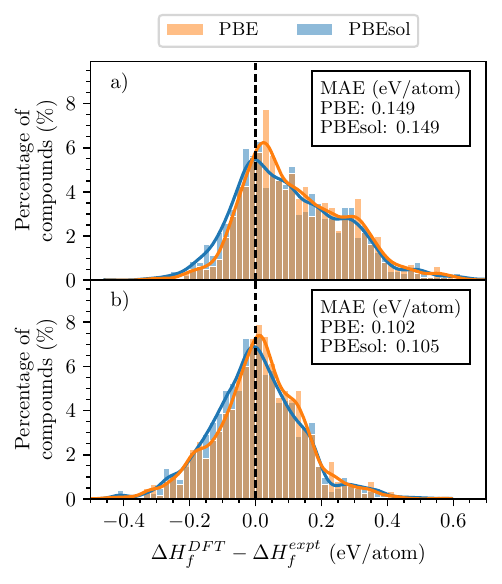}
    \caption{
    \textbf{a} Distribution of the differences between the pure DFT formation energies $\Delta H_f^{DFT}$, obtained with PBE and PBEsol, and the experimental reference formation energies $\Delta H_f^{expt}$. \textbf{b} Same as before, but with DFT formation energies corrected using the FERE-part approach.}
    \label{fig:SI-pbe_sol_expt}
\end{figure}

In the main text, we observed a small difference in the performance of the pure PBE and PBEsol formation energies. We checked this deviation and, as we show below, the difference in terms of MAE is related to a small difference in the available elemental references in the two versions of the MC3D. Indeed, when only considering structures that are present in both versions (and most importantly, considering the same elemental reference structures when calculating the formation energies), one obtains the results shown in Fig.~\ref{fig:SI-pbe_sol_expt}, where the two GGA functionals show a very similar performance with respect to the experimental references.

\clearpage

\section{MLIP vs. DFT vs. EXP}

\begin{figure}[h!]
    \centering
    \includegraphics[width=0.5\linewidth]{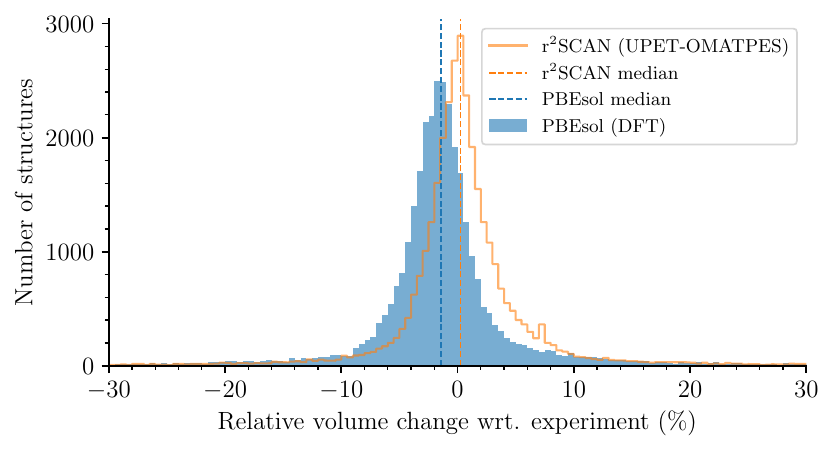}
    \caption{Relative volume change with respect to the experimental source structures for PET-OMATPES \rrscan relaxations and DFT PBEsol relaxations.}
    \label{SI-fig:r2scan-pbesol-exp-volume}
\end{figure}

Fig.~\ref{SI-fig:r2scan-pbesol-exp-volume} shows the relative volume change with respect to the initial experimental structure for the DFT results from the PBEsol-v1 version of MC3D and the PET-OMATPES relaxed version based on \rrscan. As one can notice, the MLIP relaxations exhibit a marginally better agreement with the initial experimental structures than the DFT relaxed ones using the PBEsol functional. This further highlights that the fMLIPs trained at the meta-GGA level can reach a precision and accuracy that might motivate an actual drop-in replacement of DFT with MLIPs in certain materials-science workflows. Nonetheless, especially in less explored regions of the structural space and for more advanced properties, DFT still remains the method of choice.

\clearpage

\begin{figure}[h]
    \centering
    \includegraphics[width=\linewidth]{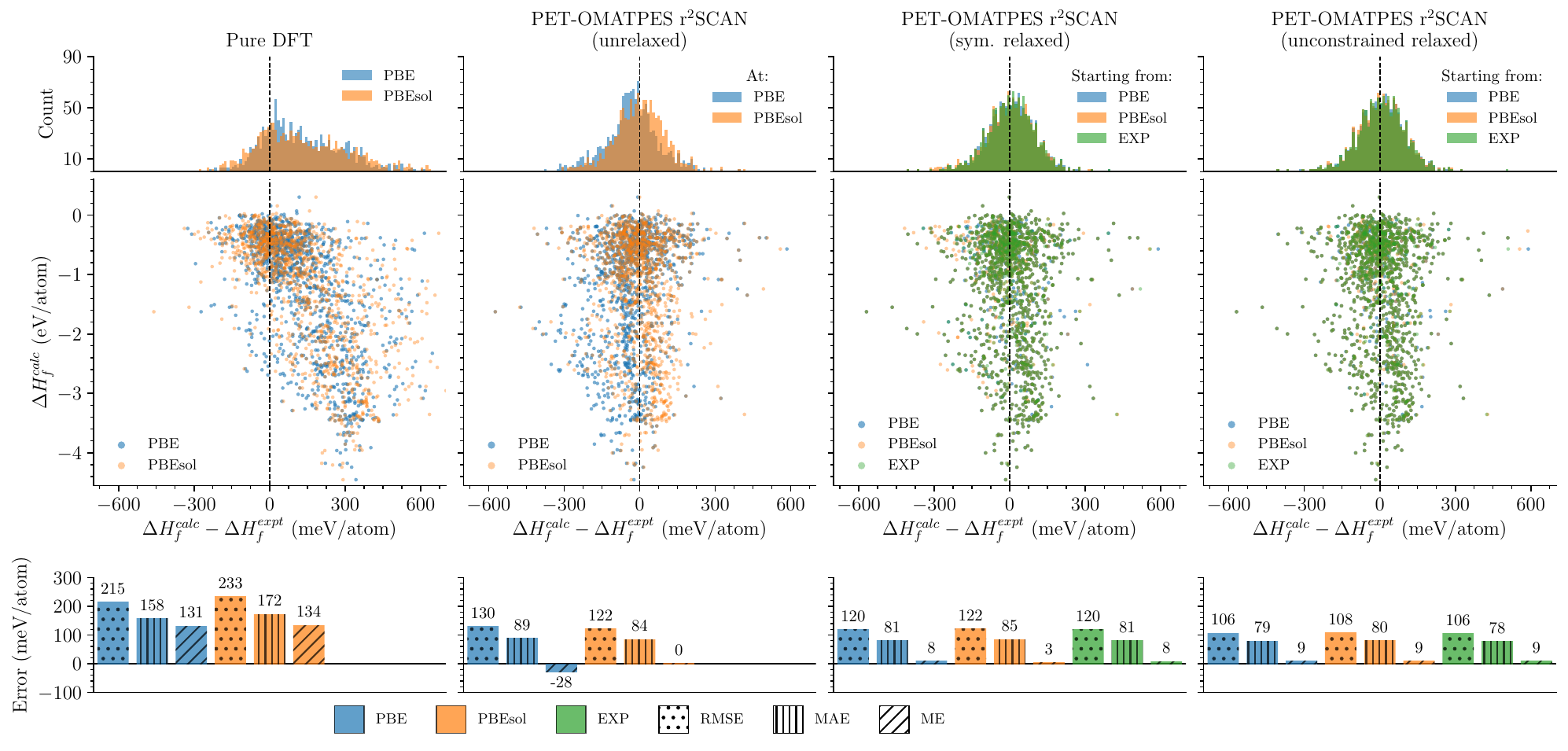}
    \caption{
    Comparison of formation energy differences with respect to experiment for several approaches based on PBE, PBEsol and the underlying experimental reference structures EXP. The different approaches ``calc'' to calculate formation energies are ``Pure DFT'' calculations \textbf{(first column)}, MLIP predicted formation energies evaluated on DFT-relaxed structures ``PET-OMATPES \rrscan (unrelaxed)'' \textbf{(second column)} and fully MLIP relaxed structures with corresponding MLIP formation energies, symmetry constrained ``PET-OMATPES \rrscan (sym. relaxed)'' \textbf{(third column)} and unconstrained ``PET-OMATPES \rrscan (unconstrained relaxed)'' \textbf{(fourth column)}, using the PET-OMATPES \rrscan MLIP. \textbf{(Top row)} Histograms of the per-structure formation energy difference $\Delta H_f^\mathrm{calc} - \Delta H_f^\mathrm{expt}$ at/starting from the PBE relaxed structures (blue), PBEsol relaxed structures (orange) and the experimental source structures EXP (green). EXP is only used for the PET-OMATPES relaxations to investigate to which extent the MLIP relaxations reproduce the ones when starting from the DFT relaxed structures. \textbf{(Middle row)} Scatter plots of the calculated formation energy $\Delta H_f^\mathrm{calc}$ against the same difference with respect to experiment, with each point representing one structure. \textbf{(Bottom row)} Summary error metrics of the differences $\Delta H_f^\mathrm{calc} - \Delta H_f^\mathrm{expt}$ with respect to experiment: root mean squared error (RMSE), mean absolute error (MAE), and mean error (ME), for PBE, PBEsol and EXP separately.
    }
    \label{SI-fig:DFT-vs-EXP-functionals}
\end{figure}

Fig.~\ref{SI-fig:DFT-vs-EXP-functionals} extends Fig.~2 in the main text by also presenting the results for unconstrained relaxations using PET-OMATPES. Even though the error metrics with respect to the experimental formation enthalpies can be further reduced compared to the symmetry constrained relaxation, especially in the context of an unconstrained MLIP architecture (such as the PET architecture~\cite{pozdnyakov_smooth_2024}) the symmetry constrained relaxation is likely the safer option~\cite{bigi_pushing_2026} for high-throughput applications where it is not possible to verify each structure.

\section{Machine learning of formation energies}

\subsection{Low dimensional representation of the structure-corrections relation}
\label{SI-ssec:ML-kpcovr}

\begin{figure}[h!]
    \centering
    \includegraphics[width=\linewidth]{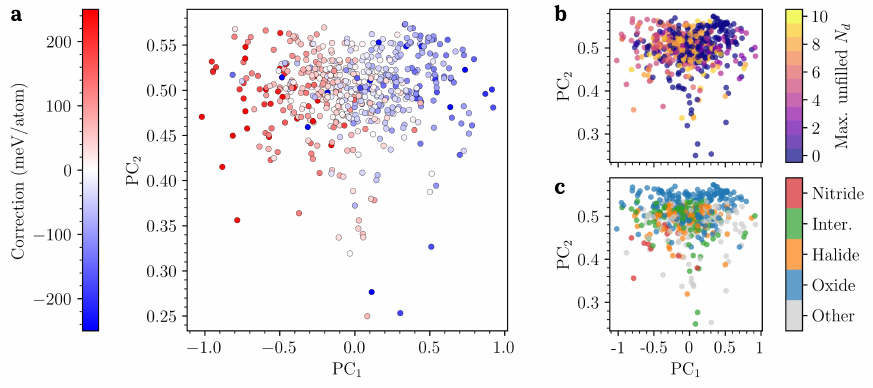}
    \caption{%
    Kernel PCovR map of the MC3D structures projected onto two principal components. The kernel PCovR uses a Laplacian kernel (consistent with the model selection in the main text) with a mixing parameter $\alpha=0.5$ optimized to balance variance preservation (PCA-like) and alignment with the formation energy correction (KRR-like). \textbf{a} Structures colored by the formation energy correction $\Delta H_f^\mathrm{corr}$. \textbf{b} The same projection colored by the maximum number of unfilled $d$ electrons $N_d$ across the elements in each compound. \textbf{c} The same projection colored by material family (nitride, intermetallic, halide, oxide, and other).
    }
    \label{fig:kpcovr}
\end{figure}

Here, we try to gain some further insights into the learned corrections. To this end, we train a Kernel Principal Covariates Regression (KPCovR) model~\cite{helfrecht_structure-property_2020}, using the same hyperparameters for the KRR model discussed before. The only difference: here we use a 50/50 train test split, to increase the sample size for the actual analysis.
The idea of KPCovR model is to combine the advantages of a low dimensional representation, like PCA, with the actual regression of a KRR model. By optimizing the combined loss, one might be able to obtain a low-dimensional representation that shows the structural variance but also an ideally high correlation of the principal components with the regression target. Fig.~\ref{fig:kpcovr}a shows a 2D PCovR map of the test set structures, whereby the color code indicates the learned correction. A clear correlation between the first principal component $PC_1$ and the correction value can be observed, with high (low) $PC_1$ values corresponding to lower (higher) corrections.

Moreover, since we start from the PET-OMATPES estimated formation energies, less systematic biases that could be related to certain anionic features are observed. Nonetheless, one can observe some correlation between simple features and $PC_1$: For example, as shown in Fig.~\ref{fig:kpcovr}b, the maximum number of unfilled $d$ valence orbitals across the elements of a given structure shows some clustering in the low dimensional representation. Almost all structures with no $d$ valence orbitals cluster in the positive region of $PC_1$, corresponding to corrections $\delta_{corr} \leq 0$. Similarly, the structures with the highest values of unfilled $d$ orbitals cluster around intermediate values ($-0.3 \leq PC_1 \leq 0.2$) and the intermediate numbers of unfilled $d$ valence orbitals tend to get the largest positive corrections ($PC_1 \leq -0.3$).

While $PC_1$ is highly correlated to the regression target, $PC_2$ mostly captures the structural similarity incorporated in the latent features of the PET model which were used as input features. Fig.~\ref{fig:kpcovr}c shows the 2D map colored by different material groups (nitrides, intermetallics (Inter.), halides and oxides). As one can see, while most of the families almost cover the full range of $PC_1$, i.e., the materials class does not directly correlate with a certain range of corrections, a clustering in the $PC_2$ dimension can be observed. 
Despite the overlap of the halides and intermetallics, which becomes slightly more separated when projecting more structures, one can clearly see the clustering of the oxides between $0.6 \leq PC_2$ and also the nitrides (which are underrepresented in this subset) that cluster in the upper left region of the map, i.e. obtaining positive corrections.

\subsection{PBEsol vs. SCAN flip rate reference based on Alexandria}

\begin{figure}[h!]
    \centering
    \includegraphics[width=\linewidth]{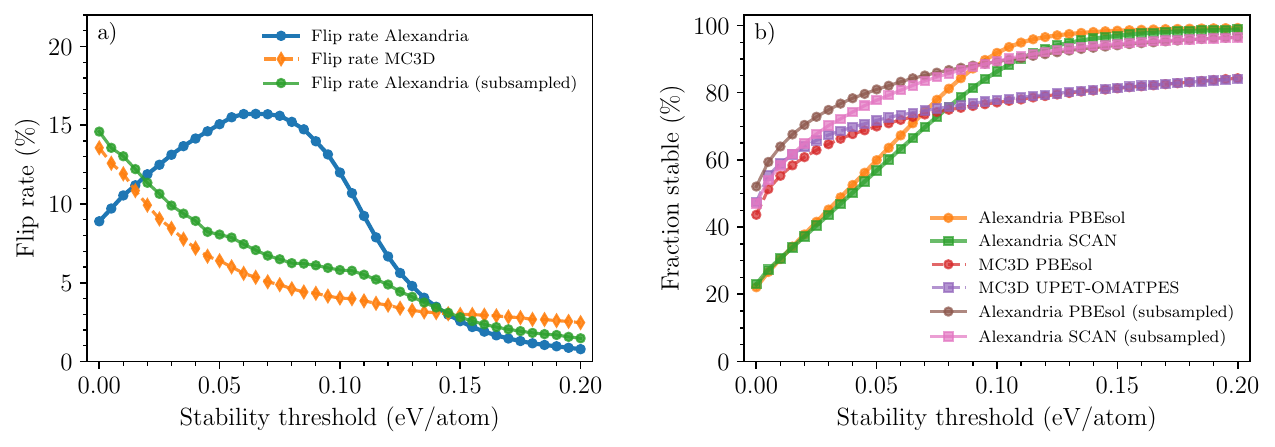}
    \caption{%
    \textbf{a)} Stability flip rate comparing DFT formation energies based on PBEsol and meta-GGA (SCAN for Alexandria and \rrscan calculated with PET-OMATPES for MC3D). The stability flip rate is the fraction of structures for which the stability classification changes between the two formation energy sets. The flip rate is shown as a function of the stability threshold, i.e. materials with an energy above hull below (above) this value are classified as stable (unstable). We compare the flip rates for the Alexandria database, the MC3D (see main text) and a subsampled version of the Alexandria database that follows the distribution of the energies above hull of the MC3D (from panel b). \textbf{b)} The fraction of stable structures as a function of the stability threshold for the PBEsol and meta-GGA version of each of the databases.
    }
    \label{fig:SI-alexandria-flip}
\end{figure}

\begin{figure}[h]
    \centering    \includegraphics[width=.5\linewidth]{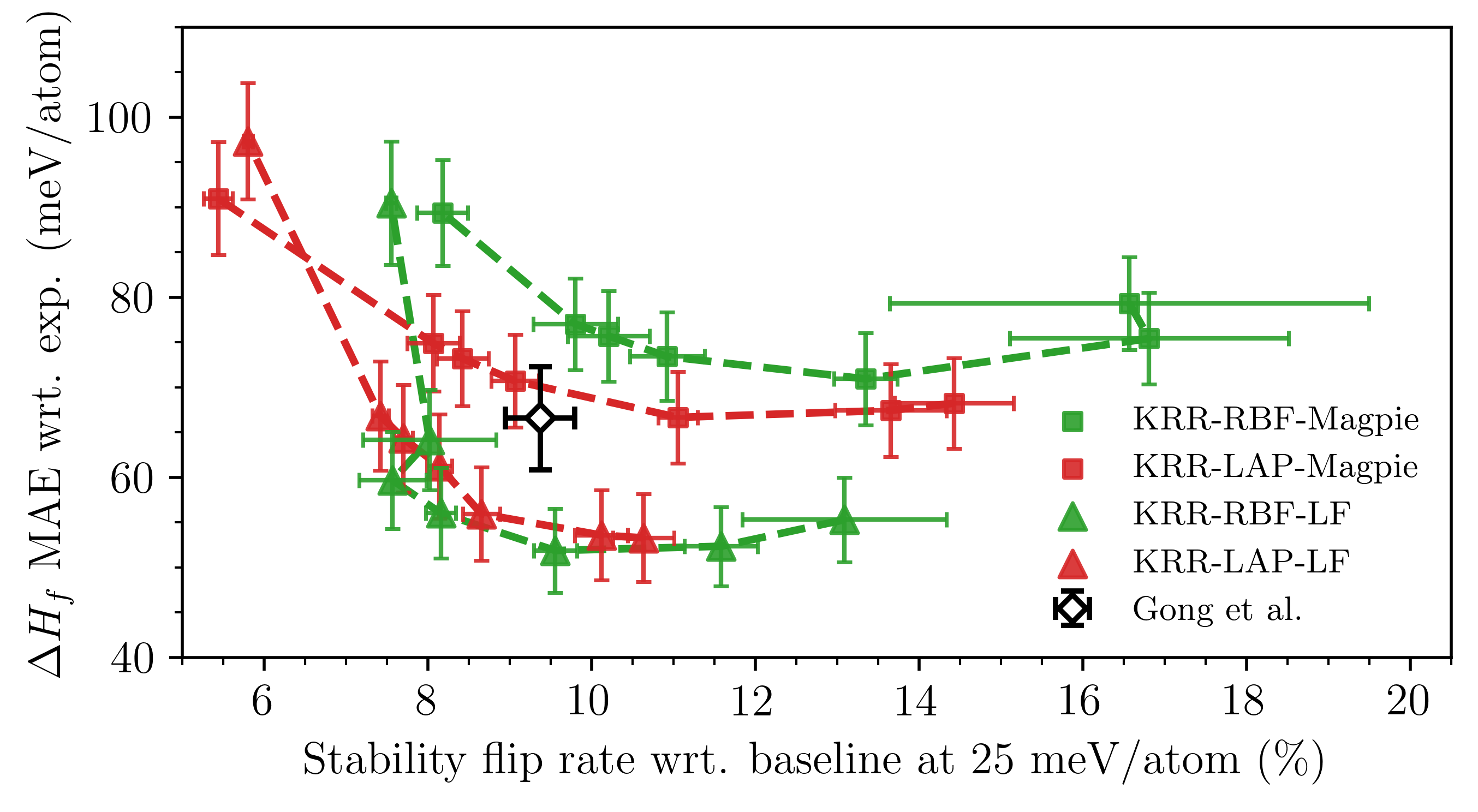}
    \caption{
    Trade-off between formation energy prediction accuracy and stability reliability for KRR models as a function of regularization strength. Each point corresponds to a specific regularization parameter $\alpha$, with dashed lines connecting points of the same model--feature combination across increasing $\alpha$ values $\alpha \in \{ 0.001, 0.01, 0.1, 0.4, 0.7, 1.0, 10\}$ (generally from left to right). The x-axis shows the stability flip rate (fraction of structures whose stability classification changes upon correction) and the y-axis the mean test MAE across 30 train-test splits (error bars: standard deviation). Results are shown for four model--feature combinations: KRR with Laplacian and RBF kernels, each paired with Magpie or LF features. In contrast to Fig.~4a in the main text, this one is based one PBEsol data. The black diamond shows the performance when retraining the model from reference~\cite{gong_calibrating_2022} on the data used in this study.
    }
    \label{SI-fig:FE-vs-flip-PBEsol-with-ref}
\end{figure}

Fig.~\ref{fig:SI-alexandria-flip} presents the energy-above-hull flip rate obtained when comparing DFT results at the PBEsol and meta-GGA (in this case \texttt{SCAN}) level. The data is based on the Alexandria database and taken from reference~\cite{schmidt_dataset_2022}, since there is currently no meta-GGA version of the MC3D available.

Fig.~\ref{fig:SI-alexandria-flip}a compares the flip rate for the 175k materials from the Alexandria database and the one discussed in the main text, when using PET-OMATPES to estimate the formation energies at the \rrscan level based on a MLIP on top of the PBEsol relaxed structures. One observes a clear difference in the trend, the Alexandria flip rates peaks at 75~meV/atom whereas the MC3D one peaks at 0~meV/atom. However, as shown in Fig.~\ref{fig:SI-alexandria-flip}b, this can mostly be attributed to the differences in the underlying distribution of materials. If the Alexandria dataset is subsampled so that the energy-above-hull values with PBEsol follow a similar distribution as the ones in MC3D, one can see that the flip rates become very similar to the ones obtained for MC3D and PET-OMATPES.

Even though the results are not fully compatible in terms of functional/method (the Alexandria data provides a PBEsol and SCAN version, whereas for MC3D we have a PBEsol version and the MLIP-based \rrscan version), the main purpose of this analysis is to confirm the general validity of the MLIP estimations beyond comparison against experimental references. Moreover, the good agreement of the flip rates justifies that the MLIP \rrscan formation energies can indeed be used as a reference for the flip rate when analyzing the ML corrections discussed in section \textbf{Balance accurate formation energies with distortion
of relative phase stability} of the main text, as the MLIP-introduced flips follow the same trend that would be expected from DFT.

\subsection{Learning from PBEsol}

\subsection{Variation of corrections across data splits}
\label{SI-ssec:correction-variation}

\begin{figure}[h]
    \centering    \includegraphics[width=.6\linewidth]{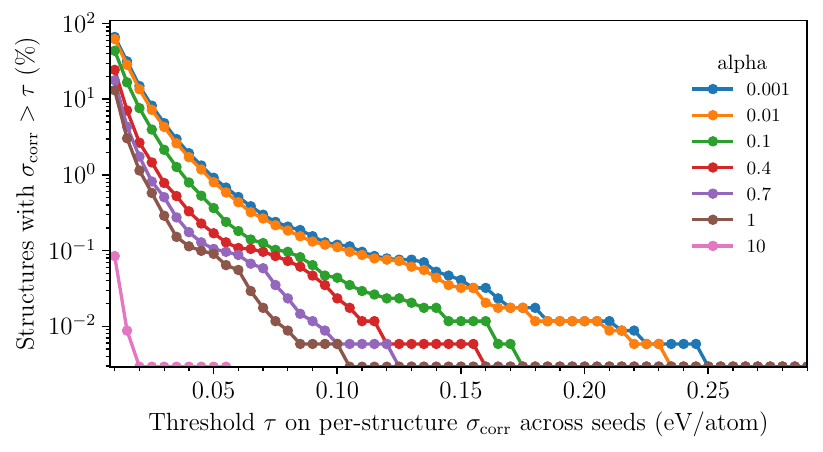}
    \caption{%
    Fraction of structures that exhibit a standard deviation of the predicted corrections $\sigma_{corr}$ that is above a certain threshold $\tau$. Here, we only focus on the KRR-LAP-LF model and the color distinguishes different regularization strengths $\alpha$.
    }
    \label{fig:SI-correction-spread}
\end{figure}

Fig.~\ref{fig:SI-correction-spread} presents the spread of the corrections across the 30 train-test splits of the dataset by showing the fraction of structures that exhibit a standard deviation of the corrections that is above a given threshold. One observes that only few cases obtain significantly different corrections based on the dataset split. As shown in the figure, these artifacts due to the overfitting of the formation energies can be significantly reduced with an increased regularization strength.

\begin{figure}[h]
    \centering    \includegraphics[width=.5\linewidth]{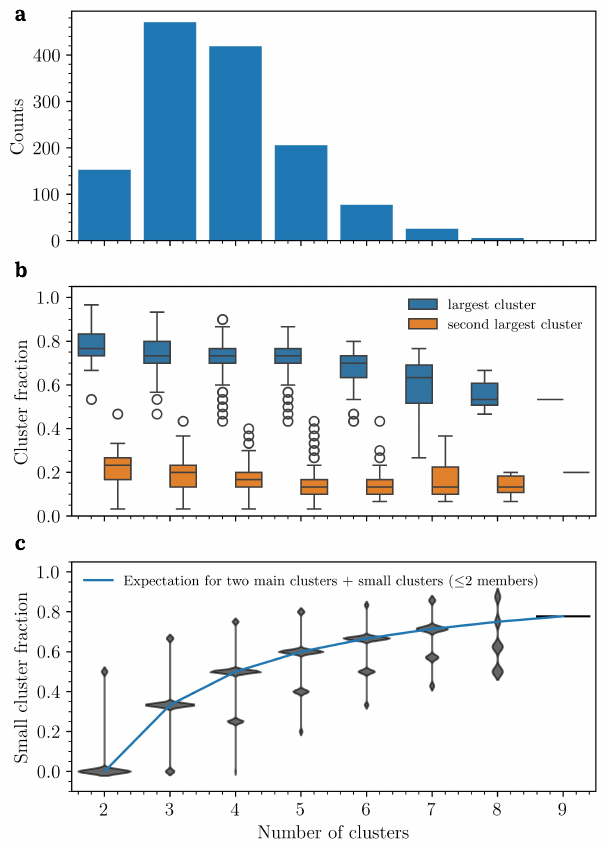}
    \caption{%
    \textbf{a)} Distribution of the number of clusters that can be found in the corrections across the 30 data splits. Only structures with a standard deviations across the corrections of more than 0.025~eV/atom are included. A cluster is determined based on a greedy single-linkage approach, i.e., corrections with a difference smaller than 10~meV/atom to the previous member (starting from a sorted list) are added to the same cluster.
    \textbf{b)} Fraction of the largest and second largest cluster grouped by the number of clusters.
    \textbf{c)} Fraction of small clusters, i.e. those with only one or two members, for each number of clusters. The blue line shows the expected fraction when assuming that the corrections are grouped into two main clusters and additional small ones.
    }
    \label{fig:SI-corrections-clustered}
\end{figure}

While the vast majority of structures obtains corrections that are consistent across the different train-test splits, there is still a small fraction that show significant variations across the different data splits. On the one hand, this could be an indication of overfitting and on the other hand it might be related to specific outliers in the training set, due to the small amount of experimental reference data (which is of course related to the overfitting).
In Fig.~\ref{fig:SI-corrections-clustered}, we further investigate how the corrections vary across the data splits. First of all, one observes in Fig.~\ref{fig:SI-corrections-clustered}a that for the majority of structures, the corrections cluster in only 3 or 4 clusters. Moreover, as shown in Fig.~\ref{fig:SI-corrections-clustered}b, the largest cluster contains around 80\% of the corrections in case of 3 or 4 clusters. Even for cases in which the corrections are distributed over multiple clusters, the large majority is still covered by the largest and second largest cluster. This is further supported by Fig.~\ref{fig:SI-corrections-clustered}c, which presents the fraction of clusters with only one or two members. In addition to the distribution shown as a violin plot, we also plot the expected fraction when assuming that the corrections are part of two major clusters and all the other corrections belong to clusters with only one or two members. As can be seen in the figure, the observed data agrees very well with the expectation. 

This analysis confirms that some data splits (e.g. the presence/absence of certain structures in the training set) can lead to differing predictions. However, from a committee and majority vote perspective, the models agree largely on a certain correction. Hence, we stick to the model with $\alpha=0.1$ that had been selected based on the discussion in the main text, as the uncertainty observed in the corrections could be mitigated by a committee of models. Nonetheless, as it can be seen in Fig.~\ref{fig:SI-correction-spread}, increasing the regularization (e.g. to $\alpha=0.4$) also helps.

\clearpage

\subsection{ML corrections vs MLIP \rrscan}

\begin{figure}[h!]
    \centering
    \includegraphics[width=0.5\linewidth]{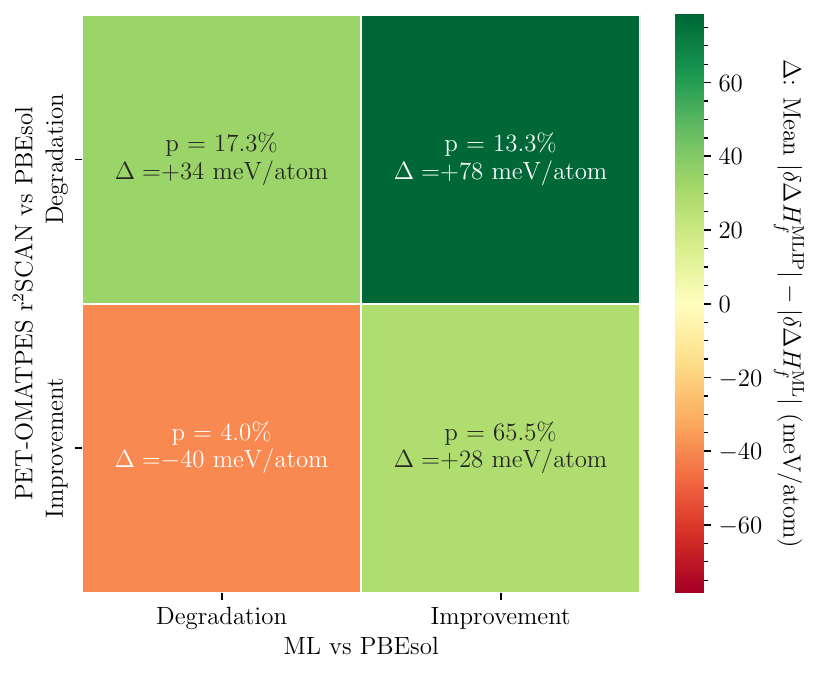}
    \caption{%
    Heatmap showing the improvement or degradation of the formation energies (with respect to experiment) relative to the PBEsol baseline for the zero-shot PET-OMATPES \rrscan formation energies, and the further ML-corrected ones. The color scale indicates the average difference in agreement with experiment between the two approaches, which is also indicated by the $\Delta$ values in each cell. Positive values correspond to a better agreement of the ML corrected formation energies with respect to experiment compared to the zero-shot PET-OMATPES \rrscan ones, i.e., $|\delta\Delta H_f^\mathrm{ML}|$ is smaller than $|\delta\Delta H_f^\mathrm{MLIP}|$. The fraction $p$ of the structures falling in each category is also indicated. The data is aggregated over the 30 test sets that were analyzed throughout this work.}
    \label{SI-fig:heatmap-improvement}
\end{figure}

Fig.~\ref{SI-fig:heatmap-improvement} shows the comparison of the improvement or degradation (with respect to experiment) of the PET-OMATPES \rrscan zero-shot formation energies and the ML-corrected ones, relative to the PBEsol baseline. In 96\% of the cases, the ML-corrected ones outperform the zero-shot formation energies. 

\begin{figure}[h!]
    \centering
    \includegraphics[width=0.6\linewidth]{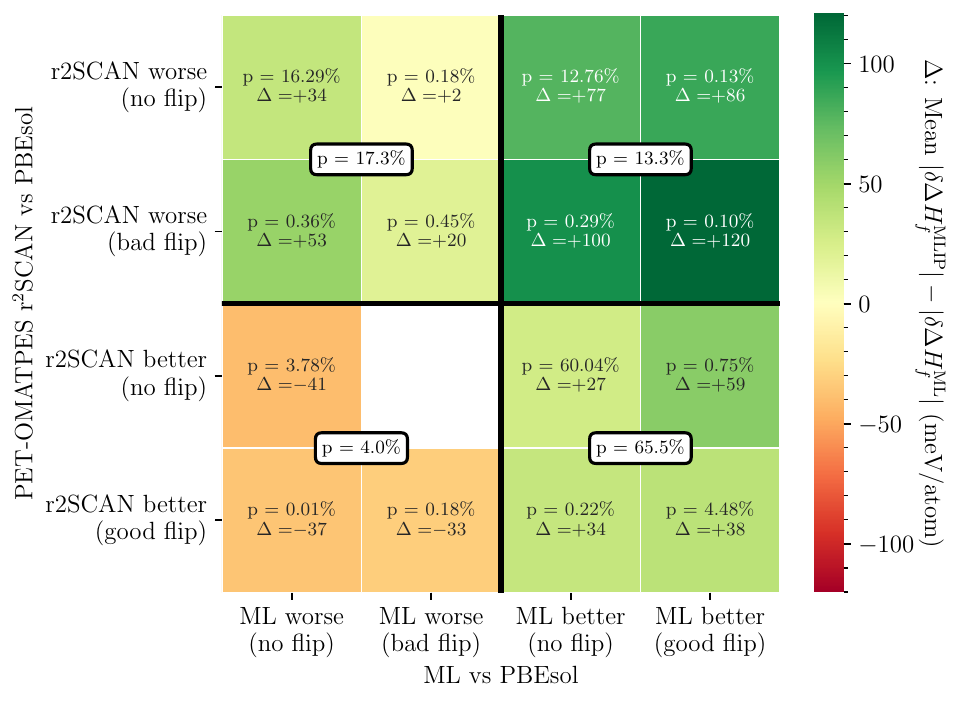}
    \caption{%
    Heatmap showing the improvement or degradation of the formation energies (with respect to experiment) relative to the PBEsol baseline for the zero-shot PET-OMATPES \rrscan formation energies, and the further ML-corrected ones. Each cell is further sub-divided according to the potential flip of the stability classification compared to the PBEsol baseline. 
    The color scale indicates the average difference in agreement with experiment between the two approaches, which is also indicated by the $\Delta$ value in each cell. Positive values correspond to a better agreement of the ML corrected formation energies with respect to experiment compared to the zero-shot PET-OMATPES \rrscan ones, i.e., $|\delta\Delta H_f^\mathrm{ML}|$ is smaller than $|\delta\Delta H_f^\mathrm{MLIP}|$. The white boxes in the center of the cells of the outer $2\times 2$ matrix report the total fraction of structures belonging to the corresponding outer group. The fraction $p$ of the structures falling in each category is also indicated. The data is aggregated over the 30 test sets that were analyzed throughout this work.
    }
    \label{SI-fig:heatmap-flips-improvement}
\end{figure}

Fig.~\ref{SI-fig:heatmap-flips-improvement} further divides the categories from Fig.~\ref{SI-fig:heatmap-improvement} into potential stability flips, essentially combining Fig.~5 in the main text with Fig.~\ref{SI-fig:heatmap-improvement}. This figure helps to further observe that the ML corrections are almost always better than the zero-shot PET-OMATPES \rrscan{} formation energies, and that most of the stability flips of the ML-corrected model (compared to PBEsol) are already induced by the physically more grounded MLIP.

\clearpage

\section{
    Comparison of FERE and ML corrections
    }
\label{SI-ssec:ML-vs-FERE}

In the following, we extend the comparison of the ML corrections against the established and widely adopted FERE-like corrections. In particular, we compare the performance of the approaches on two test sets: the test set that yields the best MAE for the ML model, and the test set (with another seed) that yields the worst performance for the ML model.

\subsection{Best test split}

\begin{figure}[h!]
  \centering
  \includegraphics[width=.9\linewidth]{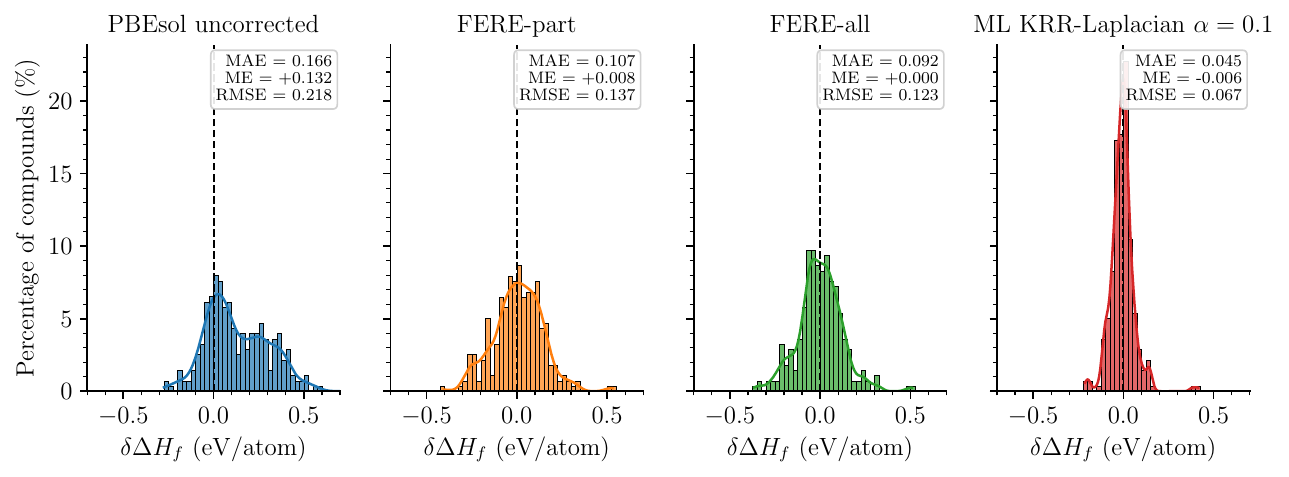}
  \caption{%
    Distributions of formation energy residuals with respect to experiment ($\delta\Delta H_f =\Delta H_f^{\mathrm{calc}} - \Delta H_f^{\mathrm{expt}}$) on the best test split ($N=277$) for
    (a)~uncorrected (pure) DFT with PBEsol,
    (b)~FERE-part,
    (c)~FERE-all, and
    (d)~ML KRR-LAP-LF (this work).
    The vertical dashed line marks zero error and the MAE and ME are annotated
    in each panel.
  }
  \label{fig:SI-best-fere-ml-error-dist}
\end{figure}

Fig.~\ref{fig:SI-best-fere-ml-error-dist} shows the distribution of errors with respect to experiment of the different methods. It is very clear that the ML corrections outperform all other methods, significantly reducing all error metrics (ME, MAE, RMSE) compared to the pure DFT and the FERE corrections. 

\begin{figure}[h!]
  \centering
  \includegraphics[width=.9\linewidth]{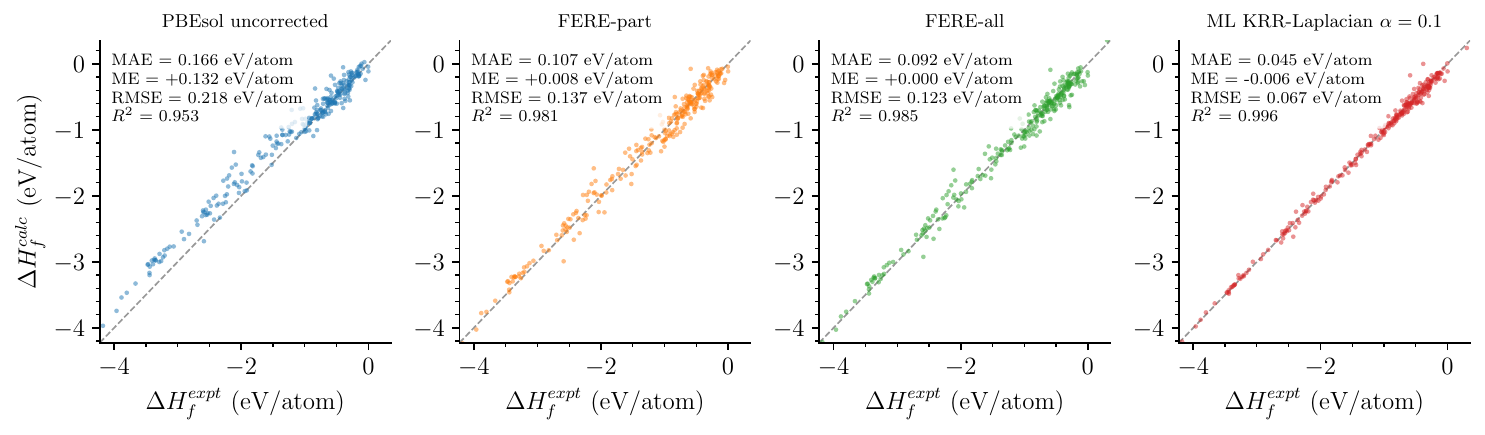}
  \caption{%
    Parity plots of calculated vs.\ experimental formation energies for the different methods on the best test split.  The diagonal represents perfect agreement.%
  }
  \label{fig:SI-best-fere-ml-parity}
\end{figure}

Fig.~\ref{fig:SI-best-fere-ml-parity} shows the parity plots of the different methods. While the FERE corrections can indeed remove the systematic bias (as also reflected in the ME in the previous figure) and also improve the correlation, they still show a noticeable spread, similar to the one of the pure DFT. Only the ML corrections can remove the systematic bias and also significantly reduce the spread.

\clearpage

\subsection{Worst test split}
\label{SI-ssec:ML-vs-FERE-worst}

\begin{figure}[h!]
  \centering
  \includegraphics[width=.9\linewidth]{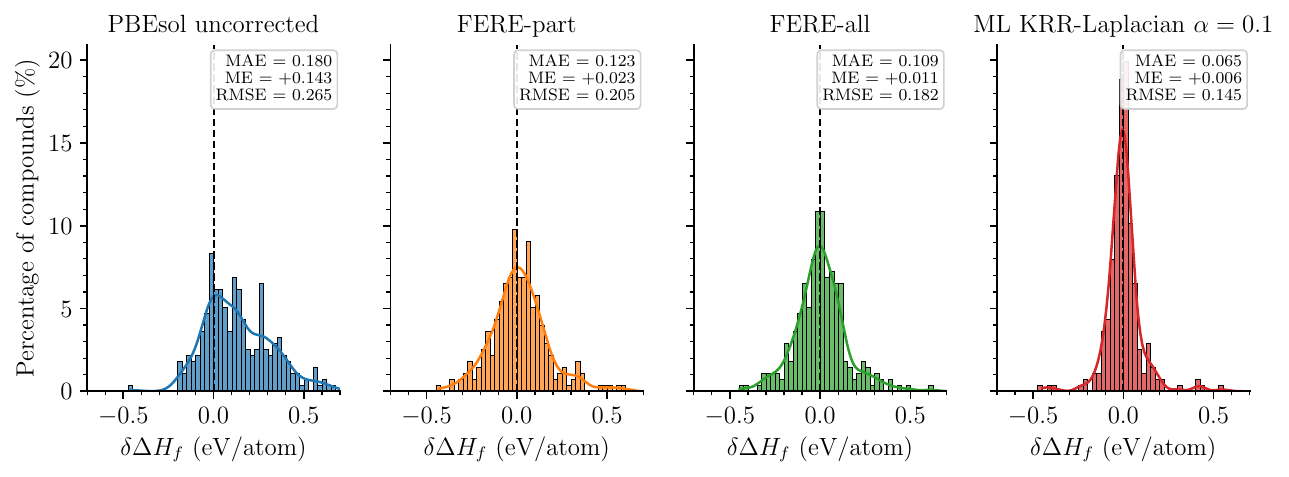}
  \caption{%
    Distributions of formation energy residuals with respect to experiment ($\delta\Delta H_f = \Delta H_f^{\mathrm{calc}} - \Delta H_f^{\mathrm{expt}}$) on the worst test split ($N=277$) for
    (a)~uncorrected (pure) DFT with PBEsol,
    (b)~FERE-part,
    (c)~FERE-all, and
    (d)~ML KRR-LAP-LF (this work).
    The vertical dashed line marks zero error and the MAE and ME are annotated in each panel.
  }
  \label{fig:SI-worst-fere-ml-error-dist}
\end{figure}

\begin{figure}[h!]
  \centering
  \includegraphics[width=.9\linewidth]{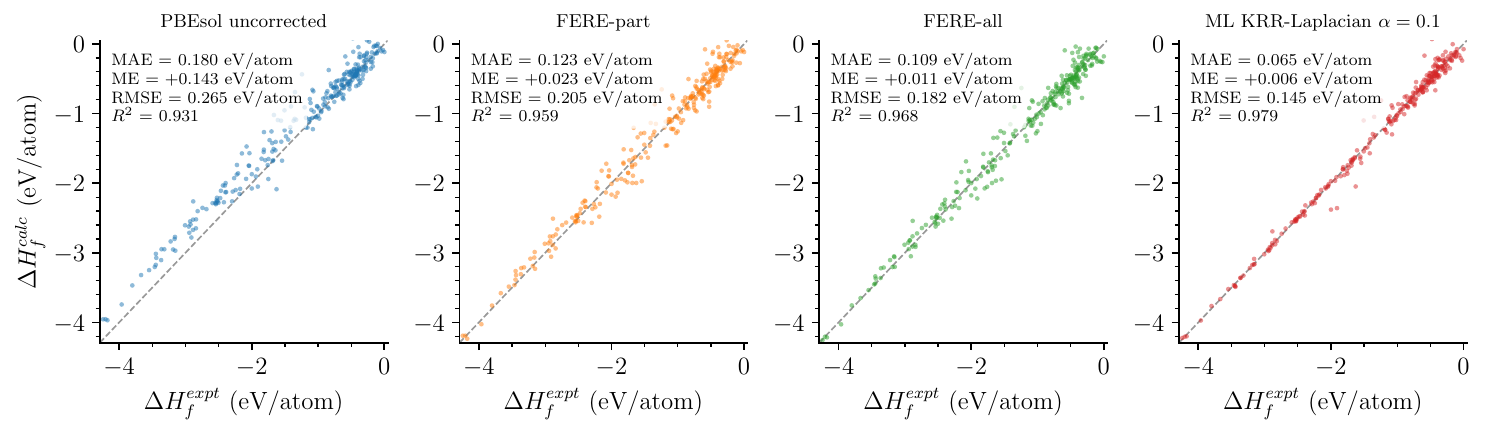}
  \caption{%
    Parity plots of calculated vs.\ experimental formation energies for the different methods on the worst test split. The diagonal represents perfect agreement.
  }
  \label{fig:SI-worst-fere-ml-parity}
\end{figure}

In this subsection, we show the same analyses as before, but this time evaluated on the test split that yields the worst MAE for the ML model. Even though the performance of the ML model is worse on this test set, it still outperforms the FERE corrections, as can be seen in Fig.~\ref{fig:SI-worst-fere-ml-error-dist} and Fig.~\ref{fig:SI-worst-fere-ml-parity}. Furthermore, even if the improvements of the error metrics are slightly reduced with respect to the previous test split, the ratios are still similar. 

It is interesting to note that the error of the pure DFT also follows the same trend. This indicates that the reduced performance is driven by certain structures that exhibit much larger errors in general. Hence, the variation is less of a model artifact, as it is also observed for DFT.

\begin{figure}[hbt]
  \centering
  \includegraphics[width=.6\linewidth]{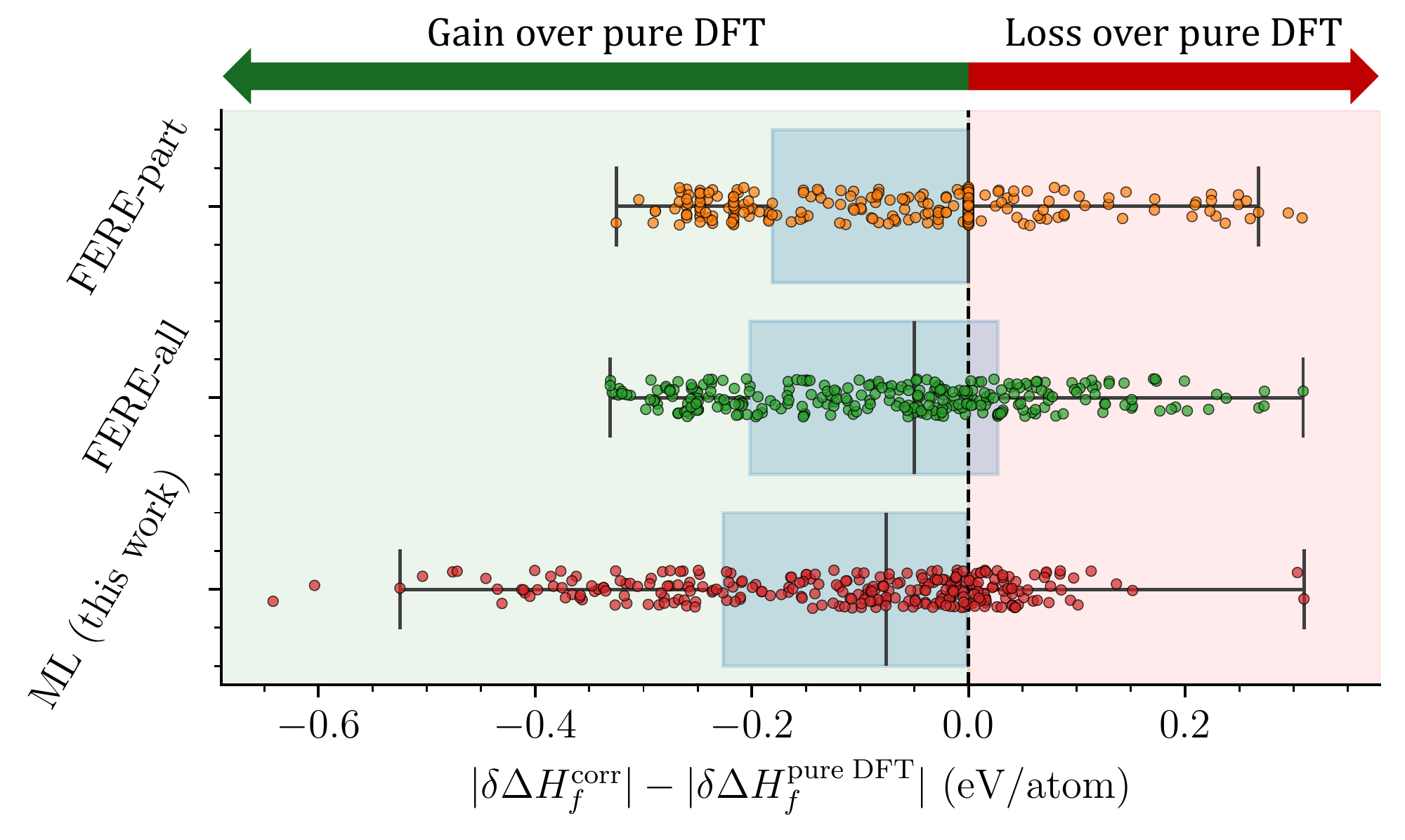}
  \caption{%
    Per-compound change in absolute deviation from experiment, $|\delta\Delta H_f^{\mathrm{corr}}| - |\delta\Delta H_f^{\mathrm{pure\, DFT}}|$, for different correction methods ``corr'': FERE-part, FERE-all, and ML (KRR-LAP-LF). Negative values (green background) indicate improvement over the pure DFT values, positive values (red background) indicate degradation. Box plots show the median (vertical line), IQR (box), and 1.5$\times$IQR whiskers; individual data points are overlaid with vertical jitter. The test set corresponding to the worst MAE for the ML model (KRR-LAP-LF) is shown here.%
  }
  \label{fig:SI-ML-FERE-delta-err}
\end{figure}

Finally, Fig.~\ref{fig:SI-ML-FERE-delta-err} shows the per-compound improvement on the test split that yields the worst performance of the ML model. This figure can be compared with Fig.~\ref{fig:ML-FERE-delta-err} in the main text, which instead presents the same analysis for the best test split. Even on this worst test split, the ML model outperforms the FERE corrections by improving the absolute error more significantly and, when it degrades the error, it does so to a smaller extent compared to the FERE corrections. 

\end{document}